\documentclass[12pt]{article}
\usepackage{mathrsfs}
\usepackage{amsfonts}
\usepackage{graphicx}
\usepackage{color}
\usepackage{amsmath, amssymb, theorem}
\usepackage{setspace,bm}
\usepackage{rotate, color}
\usepackage{graphicx, rotating}
\usepackage{natbib}

\newtheorem{theorem}{Theorem}
\newtheorem{remark}{Remark}
\newtheorem{lemma}{Lemma}

\newcommand{\ve}[1]{\bm{{#1}}}
\newcommand{\vesub}[2]{\bm{{#1}}_{#2}}
\newcommand{\vesup}[2]{\bm{{#1}}^{#2}}
\newcommand{\vess}[3]{\bm{{#1}}_{#2}^{#3}}
\newcommand{\hve}[1]{\hat{\ve{#1}}}
\newcommand{\hvesub}[2]{\hat{\ve{#1}}_{#2}}

\newcommand{\tve}[1]{\tilde{\ve{#1}}}
\newcommand{\tvesub}[2]{\tilde{\ve{#1}}_{#2}}

\newcommand{\tvess}[3]{\tilde{\ve{#1}}_{#2}^{#3}}

\begin{document}
\title{Distributed sequential method for analyzing massive data
}
\author{Zhanfeng Wang  \quad  Yuan-chin Ivan Chang\footnote{Corresponding author:
E-mail: ycchang@gate.sinica.edu.tw; ivan.chang.1@gmail.com} }
\maketitle
\baselineskip 18pt
\begin{center}
\begin{minipage}{130mm}\textbf{Abstract}:
To analyse a very large data set containing lengthy variables, we adopt a sequential estimation idea and
propose a parallel divide-and-conquer method.
We conduct several conventional sequential estimation procedures separately,
and properly integrate their results while maintaining the desired statistical properties.
Additionally, using a criterion from the statistical experiment design, we adopt an adaptive sample selection, together with an adaptive shrinkage estimation method, to simultaneously accelerate the estimation procedure and identify the effective variables.
We confirm the cogency of our methods through theoretical justifications and numerical results derived from synthesized data sets. We then apply the proposed method to three real data sets, including those pertaining to appliance energy use and particulate matter concentration.


\end{minipage}\end{center}
\section{Introduction}

While the development of modern measurement and communication technologies has frequently made data collection procedures more efficient, we as researchers have been hard-pressed to analyse and extract information from large data sets and to keep up with our data collection capacity. Although we can leverage concepts from the divide-and-conquer algorithm to analyse very large data sets---such that we can apply longstanding statistical methods without modifying existing procedures and computation facilities---the number of partitions, the size of each partition, and some tuning parameters will soon become follow-up issues. From a statistical perspective, the more important issue will be how to legitimately and effectively merge the individual result from each partition into an informative one. This is an essential issue when we apply such divide-and-conquer thinking, especially when there is a lack of consistency among the partition results.

From a computation perspective, the distributed/parallel computation method is a powerful way of accelerating the computation procedure, when statistical procedures can be `parallelized' without overly modifying their current algorithms.
Leveraging the divide-and-conquer concept, many researchers have applied distributed computation methods to statistical hypothesis testing and estimation; one can reference \citet{Chen2012, Zhang2015, Battey2015, Lu2016}, and the references therein.
These studies are proposed under fixed sample size scenarios, and there is currently a dearth of research about how to integrate sequential procedures into a distributed computation method. The current study looks to fill this research gap, at least in part.

When analysing a very large data set featuring lengthy variables,
the computation issue becomes an essential one.
Computer scientists may want to resolve these issues from algorithm and hardware perspectives, but these usually require a complicated software setup and/or modern computation facilities.
In the current study, we adopt sequential estimation methods for regression models and allow the model for each `partition' to sequentially choose from the data pool its own new subjects, until its stopping criterion is fulfilled. With a suitable selected stopping criterion, we are able to combine and integrate the models and maintain good statistical properties in such an integrated estimate.
Conventionally, how to partition a large data set into several small ones may affect the overall analytical performance.
Here, the data sizes of the individual partitions differ and depend on the corresponding performance of the model of each partition. In particular, we adopt the method of the fixed-size confidence set estimation \citep{Siegmund:1985}, such that when data recruiting is stopped, the estimates will have a prescribed accuracy. Because the accuracy of the coefficient estimates from each partition is under control, we are able to merge these estimates into one and retain the required statistical properties of the original sequential estimation.

The method of sequential analysis was established by \cite{ward1945, ward1947}, and it has been applied to many areas since then, including clinical trials, finance, engineering control, and educational and psychological tests, \textit{inter alia}. The major feature of the sequential method is that it allows the sample size to be random and depend on the observed information \citep{Chow1965, Woodroofe1982, whitehead97, Bartroff2013}.
In addition to the random sample sizes, some sequential methods allow users to recruit new observations, based on information gathered while analysing the current-stage data. This type of sequential method is common in the literature on stochastic control, educational test, active learning in machine learning, and the like \citep{Lai1982, Lord1971, Wainer2000, Deng2009}.  %
The identification of important variables is a critical feature in applications, especially where the ability to interpret model results is essential and the data set has a lengthy list of variables. To this end, we adopt the adaptive shrinkage estimate (ASE) of \cite{wangchang13}, such that we can effectively detect high-impact variables during the sequential modelling procedure.

To demonstrate our method, we conduct several sequential fixed-sized confidence set estimation procedures at once on different machines (or central processing units (CPUs)); then, once all the `sampling' procedures are stopped, we combine their estimates. This kind of computation scheme is similar to `distributed' computing in the computer science literature; it can also be viewed as a naive parallel procedure. Please note that when we already have very large data sets in hand---as in many modern data analysis scenarios---we simply acquire new observations from an existing data set and `re-estimate' the regression coefficients of the corresponding model; this situation is different from traditional sequential analysis applications, where we need to actually `collect' new samples.

The remainder of this paper is organized as follows. In Section 2, we review the general sequential fixed-sized confidence region estimate, introduce our distributed sequential estimation procedure, discuss how to combine individual estimates into a final estimate when all procedures have finished, and study the method's asymptotic properties.
Furthermore, in Section 3, we present a sequential procedure with an ASE.
Section 4 summarizes the numerical results by using the synthesized data and some real examples, including a data set pertaining to appliance energy use, and two fine particle/particulate matter (PM2.5) concentration data sets. Technical details are presented in the Appendix.

\section{Distributed sequential estimation in linear models}


Consider a linear regression model,
\begin{eqnarray}\label{lm}
	Y=\vesup {X}{\top} \ve{\beta} +\epsilon,
\end{eqnarray}
where $Y \in R$ is a response variable, $\ve{X} \in \vesup{R}{p}$ is a covariate vector with length $p$, $\ve{\beta}$ is an unknown vector of parameters to
be estimated, and $\epsilon$ denotes the random error with mean $\hbox{E}(\epsilon) =0$ and variance $\hbox{Var}(\epsilon)=\sigma^2>0$.
Building a fixed-sized confidence set for $\ve{\beta}$ with a prespecified coverage probability is a classical problem, and
some authors have extended the thinking of \cite{Chow1965} to linear regression models \cite{albert1966, gleser1965, srivastava1971}. Since then, many papers have appeared in the literature that feature various setups and perspectives \cite{GRAMBSCH1989, ChangMart1992, MuthuPoruthotage2013, Vorobeichikov:2017}.

Assume that there is already a very large data set available for analysis, similar to that seen in most `big data' scenarios.
When its size is too large---such that using the data set all at once is impractical---the divide-and-conquer method is an economical, hands-on approach that requires fewer changes to existing tools (e.g. hardware and software).
If we simultaneously run $M$ independent estimation procedures (e.g. use $M$ independent machines), then we will definitely reduce the computation time; at this point, how to partition the data set appropriately into $M$ partitions becomes an issue. Most practitioners follow the conventional sampling method by treating the available data set as a pseudo-population.
Besides those studies that examine how to divide the original data set without introducing biases, as well as partition sizes, there has been a lack of discussion in the literature regarding how to legitimately integrate all the results from each partition into a statistically meaningful estimate. This will be an important issue when there is inconsistency among the results derived from partitioning.
One of the advantages inherent in using a sequential methodology is that we do not need to specify the sample size first. For this reason, we can conduct $M$ sequential fixed-width confidence set estimation procedures on these machines. With the properties of the fixed-sized confidence set estimation procedure, the estimates of various sequential estimation procedures will have a similar estimation accuracy and coverage probabilities; this will allow us to combine them into an estimate while retaining the desired statistical properties.

\subsection{Sequential fixed-sized confidence set estimate for regression model}
Our present goal is to estimate regression parameters and build with them a confidence set of a prescribed size.
We independently run $M$ estimation procedures; each of them will independently and sequentially recruit observations from the same data set, without replacement.
When all sequential procedures are stopped, we combine their results into one estimate, such that it can retain the desired statistical properties of the conventional sequential confidence set estimation procedure with less computation time and only some of the data. We run each procedure independently, and so we first describe below an individual estimation procedure.

At stage $n$ of procedure $j$, let notation $C_{jn} = \{(y_{ji},\vesub{x}{ji}), i=1,...,n\}$ be the data set with the observations recruited from the original data set---say, $\mathcal D$.
Suppose that $\hvesub{\beta}{jn}$ is a conventional least square estimate (LSE) of $\vesub{\beta}{0}$, based on $C_{jn}$, of model \eqref{lm}; additionally, let
$$\hat\sigma^2_{jn}=\frac{1}{n-p}\sum_{i=1}^n (y_{ji}-\vess{x}{ji}{\top} \hvesub{\beta}{jn})^2$$ be an estimate of $\sigma^2$.
With the notations defined above, we now describe the $j$th sequential estimation procedure as follows.

Let $\hvesub{\beta}{jn_0}$ be an LSE based on the initial data set $C_{jn_0} =  \{(y_{ji},\vesub{x}{ji}), i=1,\ldots,n_0\}$ of size $n_0$.
Let $\mu_{jn}=\lambda_{min}[(\vesub{X}{jn}\vess{X}{jn}{\top})/n]$; $\vesub{X}{jn}$ is a design matrix based on a set of covariate vectors, $\{\vesub{x}{j1},...,\vesub{x}{jn}\}$, of the observations in $C_{jn}$.
(Notation $\lambda_{min}(A)$ denotes the minimum eigenvalue of matrix $A$.)
Let $a^2$ be the $1-\alpha$ quantile of the chi-square distribution with degree of freedom $p$ and $$\sum_{j=1}^M\tilde{a}_j^2=a^2,$$ where $\tilde{a}_j>0$.
For a given $d >0$, define a stopping time
\begin{align}\label{Ndef}
	N_j=N_{jd}\equiv\inf\left\{n\geq n_0: (\hat\sigma_{jn}^2+\frac{1}{n})\leq \frac{d^2n}{\tilde{a}_j^2\mu_{jn}}\right\}.
\end{align}
Hence, $N_j$ is the smallest number of observations, such that the inequality in \eqref{Ndef} is fulfilled.
 Let $S_{N_j}=(\ve z-\hvesub{\beta}{jN_j})^{\top}(\vesub{X}{jN_j}\vess{X}{jN_j}{\top}) (\ve z-\hvesub{\beta}{jN_j})$, and $\ve z=(z_1,\cdots, z_p)^{\top}$ and
\begin{align}\label{Rn}
	R_{N_j}\!\! =\!\left\{ \ve z \in R^p\!\!:~~\!\!\frac{S_{N_j}}{N_j}\leq \frac{d^2}{\mu_{jN_j}}\right\}.
\end{align}
When the $j$th procedure is stopped according to \eqref{Ndef}, it is shown that $R_{N_j}$ is a confidence ellipsoid of $\vesub{\beta}{0}$ with the length of a maximum axis no greater than $2d$ and the coverage probability equals approximately $1-\tilde{\alpha}_j$, where $\tilde{\alpha}_j$ is the probability corresponding to $\tilde a_j$, {which is the $1-\tilde{\alpha}_j$ quantile of $\chi^2(p)$}. The coverage probability of $R_{N_j}$ depends on both $\tilde a_j$ and $d$.
The constant $d>0$ serves as a precision index, which restricts the size of the confidence ellipsoid; this can be determined by users, based on their practical needs. When $d$ becomes smaller, we will need more observations, such that the estimate can achieve the desired precision.
There have been many discussions in the literature regarding this conventional sequential estimation procedure; hence we only briefly touch upon technical matters, in Appendix A.

{\color{black}
In a conventional sequential confidence set estimation procedure (i.e. $M=1$), we first specify the coverage, and then use its corresponding quantile and the size of the confidence set to define the stopping time. In the current method, we do not specify the coverage probability for each procedure $j$, $j=1, \ldots, M$. Instead, we only specify the coverage probability for the integrated one, which is a combination of all $M (> 1)$ procedures. Thus, $\tilde a_j$ depends on the total number of procedures $M$, and for each $j$, the coverage of $R_{N_j}$ is less than $1-\alpha$. We show that the integrated one will have the desired coverage probability. This is why we demand few observations in each procedure; it also requires less computation time.
}

\subsection{Combining several sequential estimation procedures}
Suppose we have enough computing capacity available to simultaneously support $M$ sequential estimation procedures.
When conducting a sequential procedure, we usually start with a small initial data set; technically, each procedure can have a different size of initial data set. However, to simplify our discussion, we assume that all initial data sets are of the same size ($n_0$), and we let $C_{jn_0}$, $j=1, \ldots, M$ be the initial data set for each of them.

Assume that there are already $k-1$ observations used in the estimation procedure $j$ ($k > n_0$).
If the estimates obtained from procedure $j$ based on $C_{j{k-1}}$ cannot fulfil the inequality in \eqref{Ndef}, then we will select a new observation from $\mathcal D$; from there, we will calculate the new estimates, $\hvesub{\beta}{jk}$ and $\hat\sigma^2_{jk}$, based on $C_{j{k}}$. We repeat this sequential recruiting process until the inequality in \eqref{Ndef} is satisfied.

Let $N^*=\sum_{j=1}^M N_j$ and $\rho_j=N_j/N^*$, with $j=1, \ldots, M$.
When all $M$ procedures are stopped, let
\begin{align}
	&\hve{\beta} \equiv \hvesub{\beta}{N^*}=\sum_{j=1}^M \rho_j\hvesub{\beta}{jN_j}\nonumber
\end{align}
 be a weighted average of $\{\hvesub{\beta}{jN_j}:j=1\ldots, M\}$, where $\sum_{j=1}^M \rho_j=1$.
By definition, we know that $N_j$ is a function of $d$ and $\tilde a$, which are the same for all procedures in our current setup.
Hence, $N_j$ values are random variables with the same mean (see Appendix B, \eqref{conclu3} for the proof of Theorem \ref{Th1}).
Thus, the actual number of observations used in each of the various procedures is different. Hence, we take into account the number of real observations used in each procedure, and we define $\hve{\beta}$ as the weighted average of estimates obtained from each procedure with weights equal to $\rho_j=N_j/N^*$.

Then, based on the total number of observations $N^*$, we can define a confidence set for $\vesub{\beta}{0}$ as follows:
\begin{align}\label{FRn}
  	R_{N^*}\!\! =\left\{\ve z\in R^p: (\ve z - \hve{\beta})^\top \left\{\sum_{j=1}^M \rho_j^2(\vesub{X}{jN_j}\vess{X}{jN_j}{\top})^{-1}\right\}^{-1}(\ve z - \hve\beta)\leq \frac{N^*d^2}{\mu_{N^*}}\right\},
\end{align}
where $\mu_{N^*}=\sum_{j=1}^M \rho_j\mu_{jN_j}$, $\mu_{jN_j}=\lambda_{min}[(\vesub{X}{jN_j}\vess{X}{jN_j}{\top})/N_j]$, $j=1,...,M$.
We show that $R_{N^*}$ has the desired coverage probability and estimation precision. (Detailed arguments are provided in Appendix B.)
We also prove that this distributed sequential estimation procedure retains some of the conventional statistical properties of sequential fixed-sized confidence set estimation, and summarize these properties as per the following theorem. (The proof of Theorem \ref{Th1} is in Appendix B.)
\begin{theorem}\label{Th1}
Assume that for each $j$, $\{(x_{ji}, y_{ji}), i \geq 1\}$ follows the linear regression model
(\ref{lm}), and\\
\noindent {\em (A1)} $\lim_{n\rightarrow \infty}\sum_{i=1}^{n}\vesub{x}{ji}\vess{x}{ji}{\top}/n=\Sigma$,
where the matrix $\Sigma$ is positive-definite.\\
Let $N_j$ be
defined as in (\ref{Ndef}). Then,
\begin{itemize}
\item[(1)] $\lim_{d\rightarrow 0}\frac{d^2N^*}{a^2\sigma^2\mu}=1 ~~~~\text{\rm almost surely}$, \label{Fconclu1}
\item[(2)] $\lim_{d\rightarrow 0} P(\vesub{\beta}{0} \in R_N^*)=1-\alpha$, \label{sconclu2}
\item[(3)] $\lim_{d\rightarrow 0}\frac{d^2E(N^*)}{a^2\sigma^2\mu}=1$, \label{Fconclu3}
\end{itemize}
where $\mu$ is the minimum eigenvalue of matrix $\Sigma$.
\end{theorem}
Theorems \ref{Th1} (2) and (3) state that the proposed method has the prespecified coverage probability $1-\alpha$, and that the ratio of total data used in all procedures to the `theoretical' one---the best but unknown one---is asymptotically equal to 1.
\citet{Chow1965} first named these two properties {\it asymptotic consistency} and {\it asymptotic efficiency}, respectively, to describe the asymptotic properties of a sequential estimation procedure.

Under the current setup, the conventional sequential estimation is a case with $M=1$.
The stopping time in this case is
\begin{align}\label{sNdef}
	N\equiv\inf\left\{n\geq n_0:\mbox{and}~~(\hat\sigma_n^2+\frac{1}{n})\leq \frac{d^2n}{a^2\mu_n}\right\}.
\end{align}
Obviously, $a^2>\tilde{a}_j^2$ implies that $N$ is stochastically larger than $N_j$ for each $j$, and because we run several procedures concurrently, the computation time will be smaller (as expected).
However, it is worthwhile to note that when the number of variables increases or the procedure involves variable selection, the differences in computation times will increase.
In addition, because the proposed procedure combines the estimation results of each procedure (each of which features a different data size), it is more stable than that of a single procedure in terms of coverage probabilities and the like. In our numerical studies, we will compare performance levels by using different $M$ values.

\subsection{Active learning with D-optimal design criterion}
In the machine learning literature, the term `active learning' usually refers to some learning algorithms that can interactively query users. From statistical modelling perspectives, this concept approximates subject selection. The work of \cite{Deng2009}---which extends that of \cite{JeffWu1985} and closely relates to that of \cite{robbins1951, lai1979}---gives an example of how the sequential experimental design can play a role in an active learning procedure.
In particular, \cite{Lai1982} shows that the LSE for linear regression models with adaptive designs has some nice asymptotic properties under rather general design conditions, as follows.
\begin{itemize}
\item[] {(A2)}
The random error $\{\epsilon_n\}$ is a martingale
difference sequence with respect to an increasing sequence of
$\sigma$-fields $\{{\cal{F}}_n=\sigma\{(x_j, y_j): j=1, \ldots, n\}\}$ with
\begin{align}
\sup_nE(|\epsilon_n|^\alpha |{\cal{F}}_{n-1})<\infty \mbox{ almost surely for some
}\alpha>2,\nonumber
\end{align}
\item[] {(A3)}
The maximum and minimum eigenvalues of design matrix $\sum_{i=1}^{n}x_ix_i^{T}$ satisfy,
with a probability of 1, that
\begin{align}
\lambda_{min}(\sum_{i=1}^{n}x_ix_i^{T})\rightarrow \infty \text{ and }
\log(\lambda_{max}(\sum_{i=1}^{n}x_ix_i^{T}))=o(\lambda_{min}(\sum_{i=1}^{n}x_ix_i^{T})).\nonumber 
\end{align}
\end{itemize}
These conditions are very general, and in their paper, they do not refer to any particular design.
Following their results, we will show later that if we adopt the D-optimality of statistical experimental design methods to locate new observations from the data pool, these conditions will be satisfied. Hence, with the previously defined stopping criterion, the parameter estimates will have asymptotic consistency and efficiency, as stated below.
\begin{theorem}\label{Th2}
Assume that conditions in Theorem 1, (A2), and (A3) hold.
Let $N_j$ be
defined as in (\ref{Ndef}), where new samples are selected by the aforementioned D-optimality. Then, we have
$\lim_{d\rightarrow 0}(d^2N^*)/(a^2\sigma^2\mu)=1$ almost surely, 
$\lim_{d\rightarrow 0} P(\beta_0 \in R_N^*)=1-\alpha$ and 
$\lim_{d\rightarrow 0}[d^2E(N^*)]/(a^2\sigma^2\mu)=1$, 
where {$\mu$ is the minimum eigenvalue} of matrix $\Sigma$.
\end{theorem}
Theorem \ref{Th2} simply says that with adaptive designs that use the $D$-optimality criterion, we still have asymptotic properties similar to those in Theorem 1.
(The proof of this theorem is in Appendix C.)

We know that the properties of the estimates will depend on the design method.
The D-optimal design criterion selects observations which maximize the determinant of the Fisher information matrix (or, equivalently, minimize the volume of the confidence ellipsoid for the parameters) \citep{Neyer1994, Silvey1980}.
Because the selection is based on the current estimates of the unknown parameter for each procedure,
the candidate observations for each procedure are usually different.
In addition, we adopt a `without replacement' principal, where each observation will be recruited by one procedure only.

\subsection{Computation matter}

Active learning is a special case of machine learning, and it is usually operated in a sequential way. From a statistical viewpoint, this type of method approximates a general sequential method and a statistical experimental design. When applying active learning methods under `big data' scenarios, we usually assume that the data set used in the analysis already exists. Hence, how to efficiently find informative observations sequentially, such that we can efficiently and effectively conduct the analysis, is an important research problem. In this situation, design criteria are usually used to select the promising candidates from the existing data, rather than randomly select them. The work of \cite{Deng2009} is an example.

Here we do not actually conduct a sampling process when recruiting a new observation; instead, we only draw observations from a data storage device. The sampling cost here is reduced; however, the computation cost is still noticeable, since we need to iteratively conduct the estimation process for each procedure time until its stopping criterion is fulfilled. That is, we need to compute the determinant of the matrix, repeatedly. To this end, we consider the following two computation strategies.

\subsubsection*{Iterative formulas} It is clear that in such a sequential estimation procedure, the calculations of the determinant and inverse of the design matrix demand considerable computation power. However, with help of simple algebra, both can be calculated based on their predecessors. The following properties allow us to update the determinant of the new design matrix, when we sequentially add a new observation, with simple linear algebra.
 These two formulas are common in textbooks, such as \cite{rao1973}; for convenience, we describe these below.

Let $\mathbf A$ be a nonsingular $p\times p$ matrix and $\mathbf U \in R^p$ be a column vector. Then
\begin{align}\label{inverse}
(\mathbf A + \mathbf U \mathbf U^T)^{-1} = \mathbf A ^{-1} - \frac{ (\mathbf A ^{-1} \mathbf U) ( \mathbf U^T \mathbf A ^{-1}) } {1+ \mathbf U^T \mathbf A ^{-1}\mathbf U}.
\end{align}
Moreover, we also know that
\begin{align}\label{det}
det(\mathbf A +\mathbf U \mathbf U^T) = det(\mathbf A) (1+\mathbf U^T \mathbf A^{-1} \mathbf U).
\end{align}
Equation \eqref{inverse} suggests that we can calculate the inverse of $(\mathbf A + \mathbf U \mathbf U^T)$ when we know the inverse of $\mathbf A$; Equation \eqref{det} suggests a way of calculating the determinant of $\mathbf A +\mathbf U \mathbf U^T$ when we know both the determinant and inverse of $\mathbf A$. For the sequential estimation procedures discussed here, if we are at the $k$-th stage, then $\mathbf A = \left(\sum_{i=1}^k\vesub{x}{ji}\vess{x}{ji}{\top}\right)$ and $\mathbf U = \ve{x}_{j(k+1)}$. Because we need to repeatedly calculate the inverse and determinant of the design matrix when we recruit a new observation into the procedure at each stage, these two formulas are useful, as they save us a considerable amount of computation time.

\subsubsection*{Approximation method}

Let $\rho_j=N_j/N^*$, as before; then, with a probability of 1 $$\left(\sum_{j=1}^M\frac{\rho_j^2}{N_j}\right)^{-1}\left\{\sum_{j=1}^M \rho_j^2(\vesub{X}{jN_j}\vess{X}{jN_j}{\top})^{-1} \right\}-N^*\left(\sum_{j=1}^M \vesub{X}{jN_j}\vess{X}{jN_j}{\top}\right)^{-1}\longrightarrow 0,$$
as $d$ goes to 0.
Based on this fact, we define an alternative confidence set for $\beta_0$ as follows.
\begin{align}\label{FRn1}
R^{'}_{N^*}\!\! =\left\{Z\in R^p: (\hve\beta-\ve z)^\top \left(\sum_{j=1}^M \vesub{X}{jN_j}\vess{X}{jN_j}{\top}\right)(\hve\beta-\ve z)\leq \frac{N^*d^2}{\mu_{N^*}}\right\}.
\end{align}
Obviously, it is easy to calculate \eqref{FRn1} when the number of dimensions $p$ is large, compared to (\ref{FRn}).
In addition, we show that the length of the maximum axis of the ellipsoid $R^{'}_{N^*}$ is no greater than $2d$, and that its coverage probability is approximately equal to the nominated $1-\alpha$ when the sample size is sufficiently large (or, equivalently, when $d$ is small) (see Appendix B).
Hence, besides the aforementioned iterative formulas, these asymptotic results suggest that we can use $R^{'}_{N^*}$ as an alternative confidence set, and that they provide us another way of reducing the computation cost of calculating the inverse of the high-dimensional matrix.

On the other hand, we know that
\begin{align}
	\left(\sum_{j=1}^M X_{jN_j}X_{jN_j}^{T}\right)\left\{\sum_{j=1}^M \rho_j^2(X_{jN_j}X_{jN_j}^{T})^{-1}\right\}
	&=\left(\sum_{j=1}^M \rho_j^2\right)I+\sum_{j\neq l}\rho_l^2 X_{jN_j}X_{jN_j}^{T}(X_{lN_l}X_{lN_l}^{T})^{-1} \nonumber\\
	&\geq \left(\sum_{j=1}^M \rho_j^2\right)I + \sum_{j\neq l}\rho_l \rho_j I=I,\label{I-ieq}
\end{align}
where $I$ is an identical matrix with rank $p$. This implies that
\begin{align}
	&\lambda_{max}\left\{\left(\sum_{j=1}^M X_{jN_j}X_{jN_j}^{T}\right)^{-1}\right\}\leq \lambda_{max}\left[\left\{\sum_{j=1}^M \rho_j^2(X_{jN_j}X_{jN_j}^{T})^{-1}\right\}\right].\label{comp.eigen}
\end{align}
To have the same coverage probability, the confidence set based on (\ref{FRn1}) requires more observations.
Therefore, we suggest that users use a confidence set based on (\ref{FRn1}) only for models with a lengthy parameter list, and/or when the cost of computation is high. Otherwise, we recommend using the confidence region (\ref{FRn}), especially when $p$ is not overly large.

\begin{remark}
In most `big data' analysis scenarios, data have already been collected and are stored in some digital device, such that we can easily manage and access the data: we simply `pick out' those desired observations from the available data pool, based on a predetermined selection criterion (in our method, $D$-optimality). The `sampling' cost here is lower than that in conventional situations, since there is no need to conduct new experiments or data collection processes.
\end{remark}

\section{Adaptive shrinkage estimation}\label{sec:ASE}

Model interpretation is essential in most data analysis scenarios. Users always prefer to have solutions that are interpretable and understandable. Additionally, researchers can always benefit from the interpretability of a model to validate and/or further improve their hypotheses.
By adding the ASE feature to the distributed sequential estimation procedure, we can better detect variables that significantly affect the model, and thereby increase the interpretability of the model. This is especially the case when the data set at hand contains a long list of variables.
In this section, we provide a definition for ASE, 
such that the proposed procedure can detect the effective variables for the model during its estimation process.

Suppose that only $p_0$ of $p$ components of $\vesub{\mathbf\beta}{0}$ are effective to the model,
and that we want to simultaneously identify these variables and construct a fixed-size confidence set with a prescribed coverage probability for them.
Let $\hvesub{\beta}{j}$ be the LSE of $\vesub{\beta}{0}$ in procedure $j$, as before, and
 let
$\lambda_k=\lambda |\hat{\beta}_{j}^k|^{-\gamma}$, where $\hat{\beta}_{j}^k$ is the $k$th element of $\hvesub{\beta}{j}$.
In addition, let $\lambda\equiv\lambda(n)$ be a nonrandom function of $n$, such that for some $ 0<\delta<1/2 $ and $\gamma >0$,
\begin{align}\label{cond1}
n^{1/2}\lambda \rightarrow 0 \text{ and }
n^{1/2+\gamma\delta}\lambda\longrightarrow \infty, \mbox{ as }
n\rightarrow \infty.
\end{align}
Let $\epsilon >0 $ be constant; then, define $I_{j}^k(\epsilon)=I(\sqrt n \lambda_k<\epsilon)$, for $k =1, \ldots, p$ as an indicator function, and let
\( \vesub{I}{j}(\epsilon) =\text{diag}\{I_{j}^1(\epsilon),\cdots,I_{j}^p(\epsilon)\}\) be a $p\times p$ diagonal matrix.
Then, 
$\vess{\beta}{j}{*}\equiv\vesub{I}{j}(\epsilon)\hvesub{\beta}{j}$ is an ASE of $\mathbf \beta_0$,
 %
 where ${\beta}_{j}^{*k}=0$ if $I_{j}^k(\epsilon)=0$;
 %
 otherwise, it remains the same as $\hat{\beta}_{j}^{k}$.

Suppose that $\hat\sigma_{jk}^2$ and $\hat p_0(k)=\sum_{l=1}^p I_{jl}(\epsilon)$ are estimates of $\sigma_{jk}^2$ and $p_0$, respectively, based on the observations in $C_{jk}$ at the $k$th stage.
 Let $\chi_{\hat p_0(k)}(\alpha)$ be an $1-\alpha$ quantile of the chi-square distribution with $\hat p_0(k)$ degrees of freedom, and
 let $\tilde{a}_{jk}^2=\chi_{\hat p_0(k)}(\alpha)/M$.
Then, define a stopping rule for procedure $j$ as follows:
\begin{align}\label{Ndef-var}
	\tilde{N}_j=N_d\equiv\inf\left\{k:~k\geq n_0~~\mbox{and}~~(\hat\sigma_{jk}^2+\frac{1}{k})\leq \frac{d^2k}{\tilde{a}_{jk}^2\tilde{\mu}_{jk}}\right\},
\end{align}
where $\tilde{\mu}_{jk}=\lambda_{max}[k\vesub{I}{j}(\epsilon)(\vesub{X}{jk}\vess{X}{jk}{\top})^{-1}\vesub{I}{j}(\epsilon)]$, and $\vesub{X}{jk}$ is the design matrix under the current stage.

We conduct procedure $j$ as before, until the stopping criterion $\tilde{N}_j$ is satisfied.
Please note that $\tilde{a}_{jk}^2$ in (\ref{Ndef-var}) is now random and depends on the estimated number of effective variables at the current stage. Because ${a}_{k}^2$ in (\ref{Ndef}) is a constant calculated with a total number ($p$) of variables,
this implies that $\tilde{a}_{jk}^2\leq {a}_{k}^2$ almost surely.
Therefore, ${N}_j$ is stochastically larger than $\tilde{N}_j$; this is especially the case when $p_0$ is much smaller than $p$.
These phenomena will also appear in our numerical studies.

\subsection{Adaptive shrinkage estimation in distributed sequential estimation}
When we adopt the ASE in all procedures, the major concern will be whether or not the selected variables from each procedure will be consistent.

For each $j=1, \ldots, M$, let $\tilde{N}_j$, $\vess{\beta}{j}{*}$, and $\vesub{I}{j}(\epsilon)$ be defined as before.
Let $\tilde{N}=\sum_{j=1}^M \tilde{N}_j$ be the sum of the stopping time: when all procedures are stopped, it is equal to the sum of observations used in each procedure.
Let $$\vesup{I}{*}=\prod_{j=1}^M \vesub{I}{j}(\epsilon)$$
be an element-wise product of the indicator vectors
and define the combined ASE
$$\tve{\beta}=\vesup{I}{*}(\sum_{j=1}^M \rho_j\vess{\beta}{j}{*}),$$
as a weighted average of the ASE $\vess{\beta}{j}{*}$ of each procedure
with random weights $\rho_j=\tilde{N}_j/\tilde{N}$, $j=1,\ldots, M$.
We know that for each $j$,
\begin{align}
	\sqrt{\tilde{N}_j}(\vess{\beta}{j}{*}-\vesub{\beta}{0})\longrightarrow N(0,\sigma^{2}{\vesub{I}{0}}\Sigma^{-1} {\vesub{I}{0}})
~~\text{\rm in distribution as }d\rightarrow 0,\nonumber
\end{align}
where $\vesub{I}{0}=\text{diag}\{I(\beta_{01}\neq 0),...,I(\beta_{0p}\neq 0)\}$.
Because each observation will be recruited into one procedure only, $\vess{\beta}{j}{*}, j=1,...,M$, are independent. This implies that
\begin{align}\label{Tasynorm}
  	\sqrt{\tilde{N}}(\tve{\beta}-\vesub{\beta}{0})\longrightarrow N(0,\sigma^{2}{\vesub{I}{0}}\Sigma^{-1} {\vesub{I}{0}})
~~\text{\rm in distribution as }d\rightarrow 0.
\end{align}
Hence, if $\beta_{0l}=0$, for some $1 \leq l \leq p$, $\sqrt{\tilde{N}}(\tilde{\beta}_{l}-\beta_{0l})$ will degenerate to the constant 0 when the sample size is large ($d$ tends to 0);
otherwise, it retains the same distribution as that of its corresponding component in the LSE.
Thus, Equation (\ref{Tasynorm}) implies that as $d\rightarrow 0$,
\begin{align}
	\hat\sigma^{-2}(\tve{\beta}-\vesub{\beta}{0})^\top \left[\vesup{I}{*}\left\{\sum_{j=1}^M \vesub{X}{jN_j}\vess{X}{jN_j}{\top}\right\}^{-1}\vesup{I}{*}\right]^{-}(\tve{\beta}-\vesub{\beta}{0})\longrightarrow \chi_{p_0}^2,\label{dist-var}
\end{align}
where notation $\vesup{A}{-}$ denotes a general inverse of matrix $\ve{A}$.

Let $\ve{O}$ be an orthonormal matrix, depending on the current ASE estimate, such that $(\tvesub{\beta}{1},\tvesub{\beta}{2})^{\top}=\ve{O}\tve{\beta}$ and $\ve{O}\vesup{O}{\top}=I_p$, where $\tvesub{\beta}{2}=0$ and $\tvesub{\beta}{1}$ includes all nonzero elements of $\tve{\beta}$.
We show in Appendix D that
\begin{align}
	&\hat\sigma^{-2}(\tve{\beta}-\vesub{\beta}{0})^\top \left[\vesup{I}{*}\left\{\sum_{j=1}^M \vesub{X}{jN_j}\vess{X}{jN_j}{\top}\right\}^{-1}\vesup{I}{*}\right]^{-}(\tve{\beta}-\vesub{\beta}{0})\nonumber\\
	& =\hat\sigma^{-2}(\tvesub{\beta}{1}-\vesub{\beta}{01})^\top \tvesub{\Sigma}{11}(\tvesub{\beta}{1}-\vesub{\beta}{01}),
\end{align}
where $\tvesub{\Sigma}{11}$ is defined as (\ref{matrix}) in Appendix D.
Based on this, we define a confidence set of $\tve{\beta}$ as follows.
\begin{align}\label{Rn-var}
	R_{\tilde {N}}\!\! =\!\left\{ \ve{Z} \in R^p\!\!:~~\!\!\frac{S_{\tilde {N}}}{\tilde {N}}\leq \frac{d^2}{\nu_{\tilde {N}}}\! \mbox{ and }\! z_j\!=0 \!\mbox{ for }\! I_{jj}^*\!=\!0, 1\leq j\leq p\right\},
\end{align}
where $S_{\tilde {N}}=(\ve {O}\ve{Z}-\ve{O}\tve{\beta})^{\top}\tve{\Sigma}
(\ve {O}\ve{Z}-\ve{O}\tve{\beta})$, $\nu_{\tilde {N}}=\lambda_{max}\left(\tilde {N}\vesup{I}{*}\left\{\sum_{j=1}^M \vesub{X}{jN_j}\vess{X}{jN_j}{\top}\right\}^{-1}\vesup{I}{*}\right)$ and
\begin{align}
\tve{\Sigma}=\left(
  \begin{array}{cc}
    \tvesub{\Sigma}{11}& 0 \\
    0 & 0 \\
  \end{array}
\right).\nonumber
\end{align}
Using similar arguments, it is easy to see that the maximum axis of the confidence set $R_{\tilde {N}}$ is less than $2d$, and that the distributed ASE-based sequential procedure has the following properties. (The proofs for these properties are given in Appendix D.)
\begin{theorem}\label{Th3}
Let $\tilde{N}_j$ be the stopping time defined in (\ref{Ndef-var}). Assume that the conditions of Theorem \ref{Th1} are satisfied; then,
	(1) $\lim_{d\rightarrow 0} P(\vesub{\beta}{0} \in R_{\tilde{N}})=1-\alpha$, (2) $\lim_{d\rightarrow 0}{(d^2\tilde{N})}/{(a^2\sigma^2\tilde{\mu})}=1$ \text{\rm almost surely}, and (3) $\lim_{d\rightarrow 0}{\{(d^2E(\tilde{N})\}}/{(a^2\sigma^2\tilde{\mu})}=1$,
where $a^2$ is the $1-\alpha$ quantile of $\chi_{p_0}^2$ and $\tilde{\mu}$ is the maximum eigenvalue of matrix ${\vesub{I}{0}}\Sigma^{-1} {\vesub{I}{0}}$.
Moreover, $\lim_{d\rightarrow 0}\hat p_0=p_0$ {\rm almost surely} and $\lim_{d\rightarrow 0}E(\hat p_0)=p_0$.
\end{theorem}

\begin{remark}
The divide-and-conquer method is a general concept and it can usually be applied to most procedures. However, there is no fixed sample size procedure that can concurrently guarantee the accuracy (i.e. size of confidence set) and the coverage probability of the estimate. Hence, we cannot apply the same idea to any fixed sample size procedure.
\end{remark}

\section{Numerical studies}
We apply our methods to an appliance energy consumption data set and two PM2.5 data sets collected in two major cities in China (i.e. Beijing and Shanghai). Before discussing the real-data analysis, we first use some synthesized data sets containing 500 runs for each case, to show the performance of the proposed methods and compare their results to that of the conventional sequential method (i.e. $M=1$), the latter of which we refer to as `SM'. Additionally, we offer the results of multiple parallel sequential estimation procedures ($M > 1$), which we refer to as `PSM'.

\subsection{Simulation studies}

\subsection*{Simple distributed sequential estimation}
We first consider $M=2, 5$ with $d=0.5,0.4,0.3$ and $0.2$, without using the ASE feature.
We use two different true parameter vectors:
$\vesub{\beta}{0} = (-1,1)$ and $(-1.0,1.0,0.7,0.5,0.2)$, denoted by $S_1$ and $S_2$, and generate data using the following model.
\[
	Y=\mathbf X^T\vesub{\beta}{0}+\epsilon,
\]
where {\color{black} $\mathbf X=(X_1,...,X_p)$ with $X_1=1$ and $X_i,i=2,...,p$, generated from normal distributions with mean $1$ and 0.2 for $p=2$ and $5$, and variance $1$; additionally, each component of $\epsilon$ follows the standard normal distribution.}

Table \ref{tab1-2} summarizes the total number of observations used (i.e. stopping times), empirical coverage probabilities, and computation time (i.e. CPU time).
For PSM, $N^*$ is equal to the sum of the numbers of observations used in all procedures; the computation time is the maximum computation time for all procedures.
There are two empirical coverage probabilities for PSM. The first is from the original confidence region (defined in \eqref{FRn}); the second is based on the alternative formula (defined in \eqref{FRn1}).
In this simple case, we can see from this table that the empirical coverage probabilities of all cases are close to the nominated $95\%$ coverage probability; we also see that they get closer as $d$ becomes smaller.
For a given $d$, the total numbers of observations used are similar among SM and PSMs with $M=2, 5$.
The case with a higher dimensional vector ($\vesub{\beta}{0}$) tends to use larger data sizes and the computation times are as expected. We basically see that in this simple case, the proposed method can have the same desired properties as the conventional method.

\subsection*{Using the D-optimality criterion in subject selection}
To show the procedure with D-optimality, we set only $M=2$, in addition to the conventional method.
We generate data under the same setup as in the earlier case. The purpose of this part of the simulation study is to see the performance of using the D-optimality criterion for subject selection. To mimic the aforementioned `big data' scenario, we first generate a set with $3000 \times 2$ data points, as the data pools used in the PSM procedure (i.e. 3000 data points for each of them); the SM procedure will select the joint one with $6000$ data points. We summarize the results of this study in Table \ref{tab3}.

When we compare the current results to the previous ones, we see that the procedures featuring D-optimality as its selection criterion during the recruiting process can save more computation time: in this case, the ratio of the computation time of SM to that of PSM exceeds 3\:1. This is due to the change in subject selection. The coverage is lower than the nominated $95\%$ for larger $d$ values; however, in the cases of both $S_1$ and $S_2$, they approach the nominated one as $d$ becomes smaller. The observation sizes in all cases are smaller than before; this too is due to the $D$-optimality subject selection criterion.
Please note that in conventional experiments, after we know the `theoretically best' design points, we need to collect new observations under such a specific setup; in some applications, this means conducting new experiments under some particular setup. However, in the current situation, we simply find the `best one' among the existing data points. This is often the case in `big data' scenarios, and this motivates us to develop such a method.

\subsection*{Distributed Sequential method versus divide-and-conquer}
It is known that for a linear model when the variance is unknown there is no fixed sample size solution for constructing a confidence set for regression parameters with prescribed coverage probability and precision\citep{Siegmund:1985}; this also implies that we cannot apply a simple divide and conquer (DC) method for such a problem. 
On the other hand, because the existence of nonhomogeneous data in modern large data collections is common, thus in spite of the fact that there is no solution available to the problem above  via using the thinking of divide and conquer,  we still conduct a numerical study using a contaminated  data set below to compare the performances of the proposed method under such a messy data situation in terms of estimation errors.

We generate the major part of data set using the same setup as that in the the D-optimality criterion subsection.   
In addition, we add a small portions of data generated from a  ``wrong'' regression model such that the rate of the contaminated data size to the regular data size, say $\rho$, is equal to $0.01, 0.05, 0.1$ and $0.15$.   For S1 and S2, we respectively generate two sets of noisy data using linear models with parameters $\ve{\beta}=(-5,5)$ and $(-5,5,5,5,5)$.
For the estimate baed on the divide and conquer method, we always use the whole data to estimate the regression parameters. 
Thus, the data size used in DC procedure is always equal to $6000$.  In this case, we require no data selection scheme.  

We report both the square error $SE=\sum_{i=1}^p(\hat{\beta}_i-\beta_{0i})^2$ and absolute deviation $AD=\sum_{i=1}^p|\hat{\beta}_i-\beta_{0i}|$ to assess the performances of PSM and DC.
Table \ref{tab3-1} summarizes the results of PSM and DC, where $N$ is the total sample sizes used.
Although, when $\rho=0.01$, DC has smaller SE and AD than PSM, the performance of DC decays very fast when $\rho$ increases.
Table \ref{tab3-1}  shows that SEs and ADs of PSM are much smaller than those of DC cases. 
Please note that in Table \ref{tab3-1}, when there is no or only small portion of noisy data, the PSMs use only much smaller data sizes than that of DCs, and this is the reason why PM has larger SE and AD.
However,  even in such a contaminated data situation, the PSM can locate the ``most informative'' points from a data pool via analyzing the current observations in hands, which largely reduces the chance of adding ``wrong'' data into analysis and boost the analysis efficiency.
Table \ref{tab3-2} reports the performances of DC with different $M$. 
For a total sample size $N$ in data pool, each partition has $N/M$ samples.
SE and AD (see Table \ref{tab3-2}). 
For fixed $N$, SE and AD from DC become larger when $M$ increases, especially with small $N=500$.

\begin{table*}[!h]
\caption{Simulation results with $\beta_0=(-1,1)$ and (-1.0,1.0,0.7,0.5,0.2), denoted by $S_1$ and $S_2$.}
\label{tab1-2}\tabcolsep=3pt\fontsize{7}{12}\selectfont
 \vskip 0pt
\par
\begin{center}
\begin{tabular}{cccccc c}
\hline
$\beta$&$d$ & $M$ & method & stopping time & coverage probability & computation time \\ \hline

$S_1$&0.5&1&SM&63.494(16.188)$^*$&0.95&0.086(0.026)\\
&&2&PSM&65.108(16.325)&(0.956,0.95)$^+$&0.051(0.016)\\
&&5&PSM&71.174(13.888)&(0.952,0.922)&0.027(0.01)\\
\cline{3-7}
&0.4&1&SM&99.23(19.957)&0.934&0.14(0.033)\\
&&2&PSM&99.466(22.475)&(0.932,0.924)&0.08(0.021)\\
&&5&PSM&104.104(18.914)&(0.948,0.94)&0.04(0.012)\\
\cline{3-7}
&0.3&1&SM&173.472(25.825)&0.942&0.255(0.042)\\
&&2&PSM&176.484(28.582)&(0.948,0.948)&0.141(0.028)\\
&&5&PSM&179.474(27.596)&(0.93,0.924)&0.068(0.015)\\
\cline{3-7}
&0.2&1&SM&392.276(39.207)&0.946&0.613(0.071)\\
&&2&PSM&393.452(41.269)&(0.944,0.942)&0.318(0.043)\\
&&5&PSM&400.028(39.049)&(0.956,0.954)&0.145(0.023)\\
\cline{2-7}
$S_2$&0.5&1&SM&76.964(15.167)&0.914&0.123(0.03)\\
&&2&PSM&88.196(15.5)&(0.934,0.91)&0.073(0.019)\\
&&5&PSM&113.932(14.514)&(0.938,0.882)&0.037(0.01)\\
\cline{3-7}
&0.4&1&SM&114.196(18.78)&0.934&0.192(0.039)\\
&&2&PSM&125.012(18.163)&(0.944,0.94)&0.11(0.022)\\
&&5&PSM&154.202(18.089)&(0.93,0.9)&0.055(0.013)\\
\cline{3-7}
&0.3&1&SM&192.826(25.864)&0.95&0.344(0.056)\\
&&2&PSM&205.438(22.773)&(0.942,0.93)&0.19(0.03)\\
&&5&PSM&236.398(23.574)&(0.96,0.926)&0.092(0.017)\\
\cline{3-7}
&0.2&1&SM&423.92(36.282)&0.944&0.822(0.091)\\
&&2&PSM&430.794(36.571)&(0.95,0.948)&0.42(0.052)\\
&&5&PSM&467.084(37.767)&(0.936,0.924)&0.19(0.03)\\
 \hline

 \multicolumn{7}{l}{$^{*}$ Standard deviations are in parentheses.}\\
 \multicolumn{7}{l}{$^{+}$ Coverage probabilities in parentheses are for the regions (\ref{FRn}) and (\ref{FRn1}).}
\end{tabular}
\end{center}
\end{table*}

\begin{table*}[tbp]
\caption{Simulation results of the procedures with the D-optimality criterion, to locate observations with two sets of parameters: $\beta_0=(-1,1)$ and $(-1.0,1.0,0.7,0.5,0.2)$, denoted as $S_1$ and $S_2$. The PSM results here are for $M=2$.}
\label{tab3}\tabcolsep=3pt\fontsize{7}{12}\selectfont
 \vskip 0pt
\par
\begin{center}
\begin{tabular}{cccccc}
\hline
parameter& $d$ & method & stopping time & cover probability & computation time \\ \hline
$S_1$&0.5&SM&26.108(8.119)$^*$&0.912&4.194(1.65)\\
&&PSM&27.988(6.71)&(0.932,0.928)$^+$&1.016(0.373)\\
&0.4&SM&42.702(9.816)&0.928&7.522(1.985)\\
&&PSM&42.084(9.597)&(0.936,0.936)&1.686(0.467)\\
&0.3&SM&75.554(14.305)&0.944&14.194(2.974)\\
&&PSM&75.15(14.548)&(0.938,0.938)&3.236(0.685)\\
&0.2&SM&176.196(20.177)&0.944&35.207(4.456)\\
&&PSM&176.446(19.616)&(0.958,0.958)&7.799(0.994)\\
\cline{2-6}

$S_2$&0.5&SM&47.246(11.993)&0.902&7.441(2.575)\\
&&PSM&48.86(9.748)&(0.912,0.896)&1.58(0.539)\\
&0.4&SM&72.83(12.541)&0.932&12.781(2.867)\\
&&PSM&73.794(13.367)&(0.922,0.918)&2.784(0.677)\\
&0.3&SM&132.39(16.786)&0.938&26.128(4.532)\\
&&PSM&131.668(17.604)&(0.924,0.924)&5.633(1.02)\\
&0.2&SM&299.536(25.686)&0.936&69.088(21.81)\\
&&PSM&301.656(26.456)&(0.948,0.948)&14.36(1.959)\\
 \hline
 \multicolumn{6}{l}{$^{*}$ Standard deviations are in parentheses.}\\
 \multicolumn{6}{l}{$^{+}$ Coverage probabilities in parentheses are for the regions (\ref{FRn}) and (\ref{FRn1}).}
\end{tabular}
\end{center}
\end{table*}

\begin{table*}[tbp]
\caption{Simulation results of the proposed procedure (PSM) and the common divide-and-conquer method (DC) with $M=2$, to locate observations with two sets of parameters: $\beta_0=(-1,1)$ and $(-1.0,1.0,0.7,0.5,0.2)$, denoted as $S_1$ and $S_2$.}
\label{tab3-1}\tabcolsep=3pt\fontsize{8}{12}\selectfont
 \vskip 0pt
\par
\begin{center}
\begin{tabular}{cccccc cc}
\hline
&&&\multicolumn{4}{c}{PSM}& \\
\cline{4-7}
parameter& $\rho$ & & d=0.5 &0.4 & 0.3 &0.2 & DC\\ \hline
S1&0.01&N&27.42(4.345)&38.176(8.119)&65.694(12.779)&150.636(17.864)&6000(0)\\
& &SE&0.049(0.059)&0.031(0.036)&0.017(0.022)&0.007(0.009)&0.002(0.001)\\
& &AD&0.237(0.139)&0.189(0.107)&0.136(0.081)&0.089(0.051)&0.055(0.019)\\
&0.05&N&27.636(4.459)&38.324(7.752)&66.352(11.996)&149.584(18.151)&6000(0)\\
& &SE&0.053(0.06)&0.035(0.046)&0.018(0.026)&0.008(0.01)&0.043(0.007)\\
& &AD&0.251(0.143)&0.194(0.121)&0.136(0.081)&0.093(0.052)&0.264(0.024)\\
&0.10&N&27.874(4.495)&38.378(7.892)&65.98(12.529)&150.598(19.862)&6000(0)\\
& &SE&0.053(0.057)&0.037(0.046)&0.018(0.024)&0.008(0.01)&0.171(0.016)\\
& &AD&0.253(0.139)&0.201(0.12)&0.137(0.083)&0.092(0.055)&0.53(0.029)\\
&0.15&N&27.872(4.512)&38.554(7.921)&66.194(12.023)&151.664(21.38)&6000(0)\\
& &SE&0.052(0.059)&0.033(0.041)&0.018(0.025)&0.008(0.01)&0.387(0.026)\\
& &AD&0.243(0.139)&0.188(0.114)&0.138(0.087)&0.09(0.055)&0.801(0.032)\\
S2&0.01&N&49.374(9.544)&74.348(13.115)&133.454(17.453)&301.856(26.503)&6000(0)\\
& &SE&0.061(0.049)&0.036(0.028)&0.018(0.014)&0.009(0.007)&0.003(0.002)\\
& &AD&0.424(0.156)&0.324(0.121)&0.233(0.084)&0.162(0.059)&0.093(0.032)\\
&0.05&N&48.326(9.108)&73.042(13.383)&130.872(18.231)&300.338(25.382)&6000(0)\\
& &SE&0.054(0.049)&0.033(0.028)&0.018(0.014)&0.008(0.006)&0.046(0.01)\\
& &AD&0.397(0.158)&0.312(0.122)&0.231(0.087)&0.158(0.058)&0.433(0.048)\\
&0.10&N&49.53(9.74)&74.774(13.441)&132.256(17.763)&302.33(25.535)&6000(0)\\
& &SE&0.058(0.051)&0.033(0.027)&0.018(0.014)&0.008(0.005)&0.189(0.025)\\
& &AD&0.411(0.169)&0.309(0.12)&0.229(0.08)&0.153(0.051)&0.888(0.059)\\
&0.15&N&49.112(9.981)&73.792(13.892)&132.136(18.651)&302.398(25.299)&6000(0)\\
& &SE&0.058(0.046)&0.035(0.03)&0.019(0.014)&0.008(0.006)&0.434(0.047)\\
& &AD&0.416(0.163)&0.321(0.127)&0.233(0.084)&0.159(0.057)&1.358(0.072)\\

 \hline
 \multicolumn{8}{l}{$^{*}$ Standard deviations are in parentheses.}
\end{tabular}
\end{center}
\end{table*}

\begin{table*}[tbp]
\caption{Simulation results of the common divide-and-conquer method (DC) with $M=2,5,10,15,20$ and total sample size $N=500,1000,2000$.}
\label{tab3-2}\tabcolsep=3pt\fontsize{8}{15}\selectfont
 \vskip 0pt
\par
\begin{center}
\begin{tabular}{cccccc c}
\hline
N & & M=2& 5 & 10 & 15 & 20\\
500&SE&0.043(0.013)&0.05(0.014)&0.068(0.014)&0.11(0.016)&0.248(0.025)\\
&AD&0.739(0.125)&0.797(0.114)&0.93(0.1)&1.182(0.088)&1.767(0.089)\\
1000&SE&0.02(0.006)&0.022(0.006)&0.025(0.006)&0.029(0.007)&0.034(0.007)\\
&AD&0.511(0.083)&0.529(0.08)&0.563(0.076)&0.605(0.07)&0.657(0.066)\\
2000&SE&0.01(0.003)&0.01(0.003)&0.011(0.003)&0.012(0.003)&0.012(0.003)\\
&AD&0.358(0.059)&0.364(0.058)&0.374(0.056)&0.386(0.054)&0.399(0.053)\\

 \hline
 \multicolumn{7}{l}{$^{*}$ Standard deviations are in parentheses.}
\end{tabular}
\end{center}
\end{table*}

\subsection*{Adaptive shrinkage estimation for detecting variables}

We use two vectors to show the variable-selection ability of ASE:
$\color{black}\vesub{\beta}{0}=(-2,1,1.5,2,0,0,0,0,0,0)$ and $\vesub{\beta}{0}=(-2,2,2,2,0, \ldots, 0)$.
The second vector has a total of 50 elements. Only the first four elements of this vector are nonzero; the other 46 are zero elements.
In a real-data situation, the so-called noneffective variables will not precisely equal $0$; hence, as mentioned in the previous section, we need to choose a cutting parameter, as described in Section \ref{sec:ASE} \cite[see also][]{wangchang13}.

In this simulation study, we take $M=5$ and then compare the performance of PSM to that of SM (which also has the ASE feature). In addition to the previous tables, we will report $\hat{p}_0$ here.
From Table \ref{tab4}, we first note that the estimates of $p_0$ from both SM and PSM are similar.
However, the total number of PSM observations is larger than that with SM.
This is because when there are only a few observations in the beginning of a sequential process, it is difficult to derive stable variable selection results. (PSM needs to use more observations to derive a stable estimate of the number of effective variables $p_0$.)
Despite the increase in data size in the PSM, the computation time used with PSM is around one-fifth that used with SM.
The empirical coverage probability of PSM is closer to the nominated $95\%$ level than that of $SM$.
It is worth noting that in the second case---where the true $\vesub{\beta}{0}$ has 50 elements, of which only four are nonzero---the coverage frequencies of SM are often less than $90\%$ for all $d$ values. This confirms that use of the PSM is more advantageous in high-dimensional cases.

\begin{table*}[!h]
\caption{Adaptive shrinkage estimation results with $\beta_0=(-2,1,1.5,2,0,0,0,0,0,0)$
and $(-2,2,2,2,0,...,0)$, denoted by $S_1$ and $S_2$, where there are 46 zero cases in $S_2$.}
\label{tab4}\tabcolsep=3pt\fontsize{7}{12}\selectfont
 \vskip 0pt
\par
\begin{center}
\begin{tabular}{cccccccc}
\hline
$\beta$&$d$ &  $M$& method & stopping time & coverage probability & computation time & $\hat{p}_{0}$ \\ \hline
$S_1$&0.5&1&SM&111.772(18.947)&0.898&0.155(0.035)&4.108(0.323)\\
&&5&PSM&165.812(16.988)&0.976&0.038(0.011)&3.958(0.254)\\
&0.4&1&SM&168.122(25.127)&0.92&0.267(0.053)&4.076(0.273)\\
&&5&PSM&220.206(22.532)&0.952&0.06(0.013)&3.98(0.178)\\
&0.3&1&SM&289.008(34.338)&0.91&0.577(0.105)&4.046(0.21)\\
&&5&PSM&338.978(31.237)&0.964&0.106(0.017)&3.996(0.089)\\
&0.2&1&SM&635.088(53.877)&0.934&2.414(0.484)&4.02(0.14)\\
&&5&PSM&684.934(51.767)&0.958&0.255(0.03)&4(0)\\
\cline{2-8}
$S_2$&0.5&1&SM&155.836(21.272)&0.844&0.837(0.204)&4.25(0.587)\\
&&5&PSM&358.108(17.462)&0.926&0.178(0.056)&3.992(0.089)\\
&0.4&1&SM&215.852(28.16)&0.816&1.45(0.307)&4.294(0.583)\\
&&5&PSM&413.196(27.578)&0.932&0.298(0.08)&4(0)\\
&0.3&1&SM&340.644(35.802)&0.848&2.989(0.507)&4.194(0.444)\\
&&5&PSM&540.934(36.577)&0.96&0.573(0.11)&4(0)\\
&0.2&1&SM&696.362(65.537)&0.86&10.419(2)&4.144(0.379)\\
&&5&PSM&912.79(52.673)&0.956&1.398(0.198)&4(0)\\

 \hline
 \multicolumn{8}{l}{$^{*}$ Standard deviations are in parentheses.}\\
 \multicolumn{8}{l}{$^{+}$ Coverage probabilities in parentheses are for the regions (\ref{FRn}) and (\ref{FRn1}).}
\end{tabular}
\end{center}
\end{table*}

\subsection{Real-data examples}
We apply the proposed method to appliance energy use (Energy) data and PM2.5 data collected in Shanghai and Beijing, China. For simplicity, when we apply the proposed method to these data sets, we first randomly permute the whole data set and then divide it into $M$ partitions. In this analysis, we set $M=5$---that is, we will simultaneously conduct five sequential estimation procedures with $M$ data partitions, and we will apply the SM procedure to the data at once without partitioning. It is easy for us to undertake the partitioning process as we conduct our numerical study, and we will recruit data sequentially in the previously described manner.

\subsubsection*{Appliance energy prediction data}
Data concerning appliance energy use are reported in \cite{Luis2017}, who recorded energy use and other variables such as house temperature and humidity conditions at 10-min intervals for about 4.5 months. This data set merges the records from an automatic detecting device, some wireless sensors, and weather data downloaded from a public data set. This data set is also available from a UCI machine learning repository \citep{UCI2017}.
There are 19735 records in this data set. We refer readers to their paper \citep{Luis2017} for further details. {\color{black} We study} the relationship between appliance energy use and 12 other variables---namely, temperature in the kitchen area (T1), humidity in the kitchen area (RH1), temperature in the living room area (T2), humidity in the living room area (RH2), temperature outside the building (T3), humidity outside the building (RH3), temperature in the ironing room (T4), humidity in the ironing room (RH4), pressure (Pres), wind speed (WS), visibility (V), and dew point (DEWP).

\subsubsection*{Particulate matter 2.5 data}
Particulate matter (PM) is a general term that describes the mixture of solid particles and liquid droplets in the ambient air.
The terminology `PM2.5' refers to fine particulate matter---in other words, the mass per cubic metre of air of particles with a size (diameter) generally smaller than 2.5 micrometres ($\mu m$; 2.5 micrometres is equal to 1/400 of a millimetre). Recent studies show that long-term exposure to PM2.5 might increase age-specific mortality risk, particularly from cardiovascular causes.

The PM2.5 data sets used in the current study were collected in Beijing and Shanghai; they derive from hourly-based records from January 1, 2010 to December 31, 2015 \citep{Liang2016}. After deleting missing data values, there are 49,579 and 31,880 records remaining in the Beijing and Shanghai data sets, respectively.
 We apply the proposed methods to these two data sets while using PM2.5 concentration (ug/$m^3$) as a response variable with dew point (DEWP), humidity (HUMI), pressure (PRES), temperature (TEMP), cumulated wind speed (Iws), hourly precipitation (Prec), and cumulated precipitation (Iprec) as the model covariates.

\subsubsection*{Results}
Tables \ref{tab1-real1}--\ref{tab1-real3} contain the results from using the aforementioned real data sets with procedures that use different ways of selecting new observations during sequential estimation processes---including random and $D$-optimal criterion-based selection---with $M=1$ or $5$, and $d=0.2, 0.3, 0.4,$ or $0.5$.
Table \ref{tab1-real1} reports the numbers of observations and the computation time with different computation setups. Tables \ref{tab1-real2} and \ref{tab1-real3} state the parameter estimates for Energy and PM2.5, respectively.


When we applied our method to these real data sets, we found that both the PSM and SM methods use about the same number of observations.
From Table \ref{tab1-real2}, we see that the differences in the regression parameter estimates between SM and PSM are not statistically significant. As both SM and PSM use the same model with similar numbers of observations, these results are reasonable.

On the other hand, in these real-data cases, the PSM uses less computation time than does the SM. These results suggest that the sequential estimation procedure can readily benefit from the divide-and-conquer strategy: by putting them together, we can save a considerable amount of computation time and accelerate our data analysis (or model-fitting, in this case).

When we use the D-optimality criterion to select new observations in the estimation process, both methods use fewer observations than their random selection counterparts.
However, the differences in the computation time increase.
Both methods use similar total numbers of observations under this case; hence, each estimation process in the PSM procedure, with $M=5$, uses only around one-fifth of the observations that the SM procedure uses. Therefore, the growing differences are due to the time spent in searching for new observations close to the D-optimality criterion. It is for this reason that PSM with the D-optimality feature is more efficient than the SM method under the same selection scheme.

In Tables \ref{tab1-real5}--\ref{tab1-real7}, we report the results when we add the ASE feature to the procedures.
The total number of observations used in PSM is slightly smaller than that used in SM, and the computation time of PSM is dramatically lower than that of SM (see \ref{tab1-real5}). Moreover, we can see that PSM with the ASE feature tends to select fewer variables, which is one of the reasons why PSM requires less computation time.
From Table \ref{tab1-real5}, we found that the total amount of data used by PSM is slightly smaller than that used by SM; however, the computation time of PSM is much smaller than that of SM.
We can see that PSM with ASE functionality tends to choose fewer variables, which is one of the reasons why PSM requires less computation time.
Tables \ref {tab1-real6} and \ref {tab1-real7} contain parameter estimates for all three data sets when we apply the ASE method.
When applying the ASE method, we need to select the cutting parameters. The choice of cutting parameters should depend on actual and application needs. From a model interpretation perspective, having a parsimonious model is often advantageous, because it can provide clear information for use in further research or related applications. 

\section{Discussion and closing remarks}

In this study, we applied the divide-and-conquer method to a sequential confidence estimation procedure for linear models. Together with the adaptive shrinkage estimation (ASE) methods, we were able to decide the effective variables for the models. In addition, we also used the D-optimal design criterion for selecting informative subjects for model building, both sequentially and adaptively, as in the active learning methods found in the machine learning literature. To use this kind of method, we chose the `most informative' observations, based on analysis of the current data on hand without doing extra experiments or data collection.
The proposed sequential procedure is a synergy of many useful ideas raised separately in other studies. Our numerical results show that the proposed method can effectively detect important variables for a model that requires estimation accuracy and less computation time; additionally, its adaptive sample selection feature makes it useful in scenarios where a large, precollected data set is available. Such situations are common in modern `big data' analysis. We fully exploited the adaptive sequential sampling features, such that the proposed method performed better than conventional methods in terms of estimation accuracy and computation time, while making only a minor and flexible demand on computation facilities.
We applied the proposed method to analyse one real-world data set concerning appliance energy consumption and two real-world data sets pertaining to particulate matter 2.5. Our method can detect within the data sets the important variables (from a lengthy variable list) and estimate those detected variables at a prescribed accuracy level. The features in the proposed methods are essential in addressing such problems and will provide useful information that can inform future research and/or policymaking. It is clear that we can apply this methodology to many other sequential procedures, such as sequential methods in generalized linear models and classification problems. We will report on those results elsewhere, at a later date.

\color{blue} 

\begin{table*}[!h]
\caption{Stopping times and computation times for data sets: Energy and PM2.5 in Beijing and Shanghai. The upper and lower panels involve randomly selected samples and feature a D-optimal design.}
\label{tab1-real1}\tabcolsep=3pt\fontsize{7}{12}\selectfont
 \vskip 0pt
\par
\begin{center}
\begin{tabular}{cccc cccc cccc}
 &&&&\multicolumn{4}{c}{Random selection}&&&&\\
\hline
&&&\multicolumn{4}{c}{Stopping time}&&\multicolumn{4}{c}{Computation time}\\
\cline{4-7} \cline{9-12}
Dataset & $M$ & Method & $d=0.5$ & 0.4 & 0.3 & 0.2 && $d=0.5$ & 0.4 & 0.3 & 0.2\\
\hline
Energy&1&SM& 2053 & 3200 & 5370 & 11311 && 6 & 13 & 37 & 150 \\
&5&PSM& 1876 & 2805 & 4967 & 10829 && 1 & 2 & 3 & 11 \\
\cline{2-12}
Beijing&1&SM& 4023 & 6182 & 11329 & 25193 && 15 & 33 & 116 & 557 \\
&5&PSM& 4088 & 6233 & 10981 & 25005 && 2 & 3 & 7 & 30 \\
\cline{2-12}
Shanghai&1&SM& 7559 & 11989 & 23182 & 31880$^*$ && 59 & 146 & 475 & 856 \\
&5&PSM& 8673 & 13888 & 23835 & 31880$^*$ && 6 & 13 & 37 & 69 \\
 \hline
 &&&&\multicolumn{4}{c}{D-optimal design}&&&&\\
 \hline
&&&\multicolumn{4}{c}{Stopping time}&&\multicolumn{4}{c}{Computation time}\\
\cline{4-7} \cline{9-12}
Dataset & $M$ & Method& $d=0.5$ & 0.4 & 0.3 & 0.2 && $d=0.5$ & 0.4 & 0.3 & 0.2\\
 \hline
Energy&1&SM& 287 & 522 & 1200 & 5461 && 393 & 1134 & 4772 & 67439 \\
&5&PSM& 371 & 601 & 1153 & 5597 && 11 & 31 & 87 & 932 \\
\cline{2-12}
Beijing&1&SM& 576 & 1072 & 2656 & 9608 && 3219 & 7810 & 32572 & 320896 \\
&5&PSM& 594 & 1048 & 2572 & 9589 && 51 & 115 & 396 & 3222 \\
\cline{2-12}
Shanghai&1&SM& 2122 & 3935 & 12480 & 31880$^*$ && 13913 & 39275 & 283890 & 813693 \\
&5&PSM& 2151 & 4090 & 14192 & 31880$^*$  && 210 & 583 & 5585 & 13793 \\
 \hline

\multicolumn{12}{l}{$^{*}$ The stopping criterion is not satisfied, even when all samples are used.}
\end{tabular}
\end{center}
\end{table*}

\begin{table}[!h]
\caption{Parameter estimation for the Energy data set.}
\label{tab1-real2}\tabcolsep=2pt\fontsize{7}{9}\selectfont
 \vskip 0pt
\par
\begin{center}
\begin{tabular}{cccc cccc cc}
\hline
&\multicolumn{4}{c}{$M=1$ (SM)}&&\multicolumn{4}{c}{$M=5$ (PSM)}\\
\cline{2-5} \cline{7-10}
Para.& d=0.5& 0.4& 0.3& 0.2&& d=0.5& 0.4& 0.3& 0.2\\
T1&0.381(0.05)$^{*}$&0.336(0.04)&0.337(0.03)&0.304(0.02)&&0.299(0.051)&0.367(0.04)&0.36(0.03)&0.316(0.02)\\
RH1&0.625(0.052)&0.578(0.041)&0.568(0.031)&0.551(0.021)&&0.502(0.053)&0.562(0.042)&0.57(0.032)&0.559(0.021)\\
T2&-0.416(0.068)&-0.324(0.055)&-0.343(0.041)&-0.295(0.027)&&-0.291(0.068)&-0.385(0.054)&-0.371(0.041)&-0.314(0.027)\\
RH2&-0.491(0.059)&-0.412(0.047)&-0.398(0.036)&-0.364(0.024)&&-0.336(0.06)&-0.408(0.048)&-0.419(0.036)&-0.381(0.024)\\
T3&0.226(0.049)&0.18(0.038)&0.205(0.03)&0.167(0.02)&&0.221(0.051)&0.241(0.04)&0.188(0.03)&0.159(0.02)\\
RH3&0.123(0.037)&0.084(0.029)&0.086(0.022)&0.046(0.015)&&0.035(0.038)&0.067(0.03)&0.05(0.022)&0.043(0.015)\\
T4&-0.063(0.033)&-0.082(0.027)&-0.064(0.02)&-0.066(0.014)&&-0.078(0.035)&-0.08(0.027)&-0.077(0.02)&-0.071(0.014)\\
RH4&-0.283(0.03)&-0.296(0.024)&-0.286(0.018)&-0.254(0.012)&&-0.223(0.031)&-0.231(0.025)&-0.234(0.019)&-0.258(0.013)\\
Prec&-0.039(0.015)&-0.043(0.012)&-0.036(0.009)&-0.034(0.006)&&-0.033(0.015)&-0.023(0.012)&-0.028(0.009)&-0.035(0.006)\\
WS&0.014(0.015)&0.016(0.012)&0.024(0.01)&0.022(0.006)&&0.014(0.016)&0.025(0.013)&0.032(0.01)&0.026(0.007)\\
V&0.01(0.014)&0.013(0.011)&0.004(0.009)&0.002(0.006)&&0.016(0.015)&-0.002(0.012)&-0.007(0.009)&0.002(0.006)\\
DEWP&-0.008(0.04)&-0.024(0.032)&-0.056(0.025)&-0.069(0.017)&&-0.101(0.043)&-0.101(0.034)&-0.064(0.025)&-0.047(0.017)\\
\hline
 \multicolumn{10}{l}{$^{*}$ Standard variance estimation in parentheses.}
\end{tabular}
\end{center}
\end{table}

\begin{table*}[!h]
\caption{Parameter estimation for the Beijing PM2.5 and Shanghai PM2.5 data sets.}
\label{tab1-real3}\tabcolsep=2pt\fontsize{7}{10}\selectfont
 \vskip 0pt
\par
\begin{center}
\begin{tabular}{ccc cccc ccc}
\hline
&&&\multicolumn{5}{c}{Beijing PM2.5}&&\\

$M$$^*$&Method&$d$& DEWP& HUMI& PRES& TEMP& Iws& Prec& Iprec\\
1&SM&0.5& 0.399(0.091)$^+$& 0.176(0.053)& -0.278(0.025)& -0.589(0.075)& -0.248(0.015)& -0.002(0.012)& -0.104(0.018)\\
&&0.4& 0.393(0.073)& 0.196(0.042)& -0.247(0.021)& -0.577(0.06)& -0.236(0.012)& -0.001(0.009)& -0.113(0.016)\\
&&0.3& 0.291(0.055)& 0.27(0.032)& -0.226(0.015)& -0.485(0.045)& -0.232(0.009)& -0.012(0.008)& -0.093(0.01)\\
&&0.2& 0.291(0.036)& 0.27(0.021)& -0.211(0.01)& -0.476(0.03)& -0.234(0.006)& -0.025(0.006)& -0.085(0.006)\\
\cline{2-10}
5&PSM&0.5& 0.282(0.091)& 0.29(0.052)& -0.243(0.026)& -0.484(0.075)& -0.224(0.016)& -0.092(0.044)& -0.24(0.042)\\
&&0.4& 0.283(0.073)& 0.286(0.042)& -0.246(0.021)& -0.482(0.06)& -0.24(0.012)& -0.054(0.023)& -0.21(0.028)\\
&&0.3& 0.3(0.055)& 0.258(0.031)& -0.251(0.015)& -0.504(0.045)& -0.249(0.009)& -0.074(0.017)& -0.135(0.016)\\
&&0.2& 0.35(0.036)& 0.231(0.021)& -0.223(0.01)& -0.524(0.03)& -0.235(0.006)& -0.026(0.009)& -0.12(0.008)\\
\hline
&&&\multicolumn{5}{c}{Shanghai PM2.5}&&\\

$M$&Method&$d$& DEWP& HUMI& PRES& TEMP& Iws& Prec& Iprec\\

1&SM&0.5& -0.198(0.088)& -0.051(0.042)& -0.163(0.017)& -0.255(0.083)& -0.207(0.008)& -0.013(0.01)& -0.101(0.009)\\
&&0.4& -0.135(0.07)& -0.086(0.034)& -0.176(0.014)& -0.325(0.066)& -0.21(0.006)& -0.016(0.008)& -0.096(0.008)\\
&&0.3& -0.148(0.053)& -0.08(0.025)& -0.166(0.01)& -0.312(0.05)& -0.216(0.005)& -0.012(0.005)& -0.088(0.005)\\
&&0.2& -0.11(0.045)& -0.101(0.021)& -0.164(0.008)& -0.345(0.042)& -0.216(0.004)& -0.017(0.004)& -0.084(0.004)\\

\cline{2-10}
5&PSM&0.5& -0.209(0.088)& -0.047(0.042)& -0.145(0.016)& -0.235(0.083)& -0.213(0.007)& -0.028(0.011)& -0.101(0.01)\\
&&0.4& -0.174(0.071)& -0.072(0.033)& -0.17(0.013)& -0.294(0.066)& -0.213(0.006)& -0.003(0.006)& -0.111(0.008)\\
&&0.3& -0.128(0.053)& -0.095(0.025)& -0.173(0.01)& -0.34(0.05)& -0.216(0.005)& -0.01(0.005)& -0.093(0.005)\\
&&0.2& -0.116(0.045)& -0.098(0.021)& -0.164(0.008)& -0.34(0.042)& -0.216(0.004)& -0.016(0.004)& -0.087(0.005)\\

 \hline
 \multicolumn{9}{l}{$^{*}$ Num. stands for number of machines in parallel sequential estimation.}\\
 \multicolumn{9}{l}{$^{+}$ Standard variance estimation in parentheses.}
\end{tabular}
\end{center}
\end{table*}

\begin{table*}[!h]
\caption{Number of effective variables, and stopping and computation times, with ASE for the data sets: Energy and PM2.5 in Beijing and Shanghai.}
\label{tab1-real5}\tabcolsep=3pt\fontsize{7}{9}\selectfont
 \vskip 0pt
\par
\begin{center}
\begin{tabular}{cccc cccc cccc ccccc}
\hline
&&&\multicolumn{4}{c}{Stopping time}&&\multicolumn{4}{c}{Computation time} &&\multicolumn{4}{c}{$\hat{p}_0$}\\
\cline{4-7} \cline{9-12} \cline{14-17}
Dataset & $M$ & Method& d=0.5& 0.4& 0.3& 0.2&& d=0.5& 0.4& 0.3& 0.2&& d=0.5& 0.4& 0.3& 0.2\\
\hline
Energy&1&SM& 1680 & 2563 & 4743 & 10627 && 47 & 131 & 765 & 8583 && 9 & 10 & 11 & 12 \\
&5&PSM& 1439 & 2196 & 4001 & 8977 && 1 & 3 & 13 & 91 && 6 & 6 & 8 & 8 \\
\cline{2-17}
Beijing&1&SM& 3666 & 5735 & 10305 & 25193 && 351 & 1241 & 6722 & 92463 && 7 & 7 & 7 & 8 \\
&5&PSM& 2931 & 4147 & 10539 & 24013 && 9 & 25 & 105 & 907 && 6 & 6 & 7 & 7 \\
\cline{2-17}
Shanghai&1&SM& 7040 & 10703 & 20900 & 31880 && 2249 & 7149 & 51720 & 187753 && 7 & 7 & 7 & 8 \\
&5&PSM& 4878 & 8171 & 15355 & 28660 && 20 & 102 & 430 & 1427 && 4 & 4 & 4 & 6 \\
 \hline
\end{tabular}
\end{center}
\end{table*}

\begin{table}[!h]
\caption{Parameter estimation for the Energy data set.}
\label{tab1-real6}\tabcolsep=2pt\fontsize{7}{9}\selectfont
 \vskip 0pt
\par
\begin{center}
\begin{tabular}{cccc cccc cc}
\hline
&\multicolumn{4}{c}{$M=1$ (SM)}&&\multicolumn{4}{c}{$M=5$ (PSM)}\\
\cline{2-5} \cline{7-10}
Para.& d=0.5& 0.4& 0.3& 0.2&& d=0.5& 0.4& 0.3& 0.2\\
T1&0.376(0.055)&0.378(0.045)&0.348(0.032)&0.305(0.021)&&0(0)&0.326(0.045)&0.37(0.033)&0.322(0.023)\\
RH1&0.615(0.058)&0.608(0.046)&0.573(0.033)&0.546(0.022)&&0.518(0.062)&0.541(0.048)&0.575(0.035)&0.565(0.023)\\
T2&-0.411(0.076)&-0.376(0.061)&-0.353(0.044)&-0.291(0.028)&&0(0)&-0.346(0.06)&-0.386(0.045)&-0.321(0.03)\\
RH2&-0.48(0.066)&-0.444(0.052)&-0.398(0.038)&-0.359(0.025)&&-0.353(0.069)&0(0)&-0.414(0.039)&-0.394(0.027)\\
T3&0.227(0.055)&0.206(0.043)&0.214(0.031)&0.166(0.021)&&0.248(0.056)&0.225(0.045)&0.213(0.032)&0.151(0.022)\\
RH3&0.12(0.041)&0.096(0.033)&0.09(0.024)&0.046(0.016)&&0(0)&0(0)&0(0)&0(0)\\
T4&-0.062(0.036)&-0.085(0.029)&-0.064(0.021)&-0.073(0.014)&&0(0)&0(0)&-0.083(0.022)&-0.074(0.015)\\
RH4&-0.261(0.033)&-0.288(0.027)&-0.284(0.02)&-0.259(0.013)&&-0.225(0.034)&-0.234(0.028)&-0.23(0.02)&-0.253(0.014)\\
Prec&0(0)&-0.043(0.013)&-0.033(0.01)&-0.035(0.006)&&0(0)&0(0)&0(0)&0(0)\\
WS&0(0)&0(0)&0(0)&0.023(0.007)&&0(0)&0(0)&0(0)&0(0)\\
V&0(0)&0(0)&0(0)&0(0)&&0(0)&0(0)&0(0)&0(0)\\
DEWP&0(0)&0(0)&-0.059(0.026)&-0.065(0.017)&&-0.098(0.049)&0(0)&0(0)&0(0)\\

\hline
 \multicolumn{10}{l}{$^{*}$ Standard variance estimation in parentheses.}
\end{tabular}
\end{center}
\end{table}

\begin{table*}[!h]
\caption{Parameter estimation for the Beijing PM2.5 and Shanghai PM2.5 data sets.}
\label{tab1-real7}\tabcolsep=2pt\fontsize{7}{9}\selectfont
 \vskip 0pt
\par
\begin{center}
\begin{tabular}{ccc cccc ccc}
\hline
&&&\multicolumn{5}{c}{Beijing PM2.5}&&\\

$M$&Method&$d$& DEWP& HUMI& PRES& TEMP& Iws& Prec& Iprec\\
&SM&0.5&0.397(0.096)&0.174(0.055)&-0.275(0.027)&-0.593(0.079)&-0.248(0.016)&0(0)&-0.099(0.018)\\
&&0.4&0.401(0.077)&0.193(0.044)&-0.246(0.021)&-0.577(0.063)&-0.24(0.013)&0(0)&-0.112(0.016)\\
&&0.3&0.317(0.057)&0.251(0.033)&-0.224(0.016)&-0.507(0.047)&-0.232(0.009)&0(0)&-0.091(0.01)\\
&&0.2&0.291(0.036)&0.27(0.021)&-0.211(0.01)&-0.476(0.03)&-0.234(0.006)&-0.025(0.006)&-0.085(0.006)\\
\cline{2-10}
5&PSM&0.5&0(0)&0.354(0.062)&-0.253(0.03)&-0.41(0.09)&-0.221(0.018)&0(0)&-0.271(0.036)\\
&&0.4&0(0)&0.29(0.051)&-0.255(0.025)&-0.475(0.074)&-0.235(0.015)&0(0)&-0.237(0.032)\\
&&0.3&0.288(0.056)&0.263(0.032)&-0.25(0.016)&-0.492(0.046)&-0.25(0.009)&0(0)&-0.129(0.011)\\
&&0.2&0.346(0.037)&0.233(0.021)&-0.224(0.01)&-0.52(0.031)&-0.237(0.006)&0(0)&-0.124(0.008)\\
\hline
&&&\multicolumn{5}{c}{Shanghai PM2.5}&&\\

$M$&Method&$d$& DEWP& HUMI& PRES& TEMP& Iws& Prec& Iprec\\

1&SM&0.5&-0.222(0.092)&-0.042(0.044)&-0.165(0.017)&-0.237(0.086)&-0.208(0.008)&0(0)&-0.1(0.01)\\
&&0.4&-0.164(0.074)&-0.073(0.035)&-0.178(0.014)&-0.299(0.069)&-0.21(0.006)&0(0)&-0.107(0.009)\\
&&0.3&-0.156(0.055)&-0.079(0.026)&-0.169(0.01)&-0.306(0.052)&-0.212(0.005)&0(0)&-0.093(0.006)\\
&&0.2&-0.11(0.045)&-0.101(0.021)&-0.164(0.008)&-0.345(0.042)&-0.216(0.004)&-0.017(0.004)&-0.084(0.004)\\
\cline{2-10}
5&PSM&0.5&0(0)&0(0)&-0.133(0.021)&0(0)&-0.221(0.01)&0(0)&-0.12(0.011)\\
&&0.4&0(0)&0(0)&-0.151(0.016)&0(0)&-0.212(0.008)&0(0)&-0.104(0.009)\\
&&0.3&0(0)&0(0)&-0.166(0.012)&0(0)&-0.214(0.006)&0(0)&-0.098(0.006)\\
&&0.2&0(0)&-0.092(0.023)&-0.17(0.009)&-0.333(0.045)&-0.215(0.004)&0(0)&-0.089(0.005)\\

 \hline

 \multicolumn{9}{l}{$^{*}$ Standard variance estimation in parentheses.}
\end{tabular}
\end{center}
\end{table*}

\newpage

\color{black}
\section*{Appendix}
\setcounter{equation}{0}
\renewcommand{\theequation}{A.\arabic{equation}}


\subsection*{A. Properties of sequential estimation for the machine $j$}

At the stopping time $N_j$ defined in (\ref{Ndef}), a confidence ellipsoid of $\vesub{\beta}{0}$ is
\begin{align}
\frac{(\ve z-\hvesub{\beta}{jN_j})^{\top}(\vesub{X}{jN_j}\vess{X}{jN_j}{\top})
(\ve z-\hvesub{\beta}{jN_j})}{N_j}\leq \frac{d^2}{\mu_{jN_j}}.\label{ellip1}
\end{align}
The length of the maximum axis of this ellipsoid is
$$D=2\left(\frac{N_jd^2}{\mu_{jN_j}}\right)^{1/2}\lambda^{1/2}_{max}\left[\left(\vesub{X}{jN_j}\vess{X}{jN_j}{\top})^{-1}\right\}\right]=
2\left(\frac{d^2}{\mu_{jN_j}}\right)^{1/2}\lambda^{1/2}_{min}\left[\left(\vesub{X}{jN_j}\vess{X}{jN_j}{\top})/N_j\right\}\right]
,$$
where $\lambda_{max}(A)$ is the maximum eigenvalue of matrix $A$.
By definition, $\mu_{jN_j}=\lambda_{min}[(\vesub{X}{jN_j}\vess{X}{jN_j}{\top})/{N_j}]$. Hence, $D=2d$.
For the stopping time and confidence set for the machine $j$, we show some statistical properties as presented in the following Lemma.

\begin{lemma}\label{lm1}
	Assume that the conditions of Theorem \ref{Th1} are satisfied, and ${N}_j$ is defined in (\ref{Ndef}). Then,
\begin{align*}
&\lim_{d\rightarrow 0}\frac{d^2N_j}{\tilde{a}_j^2\sigma^2\mu}=1 ~~~~\text{\rm almost surely},  \\
&\lim_{d\rightarrow 0} P(\beta_0 \in R_{N_j})=1-\tilde{\alpha}_j, \\
&\lim_{d\rightarrow 0}\frac{d^2E(N_j)}{\tilde{a}_j^2\sigma^2\mu}=1=1,
\end{align*}
where $\tilde{\alpha}_j$ satisfies $P(\chi_p^2>\tilde{a}_j^2)=\tilde{\alpha}_j$, and $\mu$ is the minimum eigenvalue of matrix $\Sigma$.
\end{lemma}

\noindent Proof. Similar to \cite{wangchang13}, the proof of this lemma is straightforward and omitted here.

\subsection*{B. Properties of distributed sequential estimation}


Since derivatives of distributed sequential estimation with $M > 1$ machines are similar to those with $M=2$, hereafter in this section---and without loss of generality---let $M=2$.

\noindent{\bf Maximum axis of $R_{N^*}$}:
From the definition of the ellipsoid $R_{N^*}$ defined in (\ref{FRn}), the
length of the maximum axis of the set defined by
\begin{align}
(\ve z - \hve{\beta})^\top \left[\left\{\rho_1^2(\vesub{X}{1N_1}\vess{X}{1N_1}{\top})^{-1}+\rho_2^2(\vesub{X}{2N_2}\vess{X}{2N_2}{\top})^{-1}\right\}\right]^{-1}(\ve z - \hve\beta)= \frac{N^*d^2}{\mu_{N^*}},\nonumber
\end{align}
is
$$D=2\left(\frac{N^*d^2}{\mu_{N^*}}\right)^{1/2}\lambda^{1/2}_{max}\left[\left\{\rho_1^2(\vesub{X}{1N_1}\vess{X}{1N_1}{\top})^{-1}+\rho_2^2(\vesub{X}{2N_2}\vess{X}{2N_2}{\top})^{-1}\right\}\right].$$
This easily shows that
\begin{align}
&\lambda_{max}\left[\left\{\rho_1^2(\vesub{X}{1N_1}\vess{X}{1N_1}{\top})^{-1}+\rho_2^2(\vesub{X}{2N_2}\vess{X}{2N_2}{\top})^{-1}\right\}\right]\nonumber\\
& \leq
\lambda_{max}\left[\left\{\rho_1^2(\vesub{X}{1N_1}\vess{X}{1N_1}{\top})^{-1}\right\}\right]+\lambda_{max}\left[\left\{\rho_2^2(\vesub{X}{2N_2}\vess{X}{2N_2}{\top})^{-1}\right\}\right].
\nonumber
\end{align}
Hence, with $\rho_j=N_j/N^*$, we have
\begin{align}
D&
\leq
2d\left(\frac{N^*}{\mu_{N^*}}\right)^{1/2}\left\{\frac{\rho_1^2}{N_1}\lambda_{max}\left[\left\{N_1(\vesub{X}{1N_1}\vess{X}{1N_1}{\top})^{-1}\right\}\right]
+\frac{\rho_2^2}{N_2}\lambda_{max}\left[\left\{N_2(\vesub{X}{2N_2}\vess{X}{2N_2}{\top})^{-1}\right\}\right]\right\}^{1/2}\nonumber\\
&=
2d\left(\frac{N^*}{\mu_{N^*}}\right)^{1/2}\left\{\frac{\rho_1^2}{N_1}\mu_{1N_1}
+\frac{\rho_2^2}{N_2}\mu_{2N_2}\right\}^{1/2}\nonumber\\
&=2d\left(\frac{1}{\mu_{N^*}}\right)^{1/2}\left\{\rho_1\mu_{N_1}
+\rho_2\mu_{N_2}\right\}^{1/2}
=2d.\label{precision}
\end{align}

\noindent{\bf Proof of Theorem 1:} At the stopping time $N_j$, from Lemma 4 and for the confidence set $R_{N_j}$ defined in (\ref{Rn}), we have
\begin{align}
&\lim_{d\rightarrow 0}\frac{d^2N_j}{\tilde{a}_j^2\sigma^2\mu}=1 ~~~~\text{\rm almost surely}, \label{conclu1} \\
&\lim_{d\rightarrow 0} P(\beta_0 \in R_{N_j})=1-\tilde{\alpha}_j, \label{conclu2}\\
&\lim_{d\rightarrow 0}\frac{d^2E(N_j)}{\tilde{a}_j^2\sigma^2\mu}=1, \label{conclu3}
\end{align}
where $\tilde{\alpha}_j$ satisfies $P(\chi_p^2>\tilde{a}_j^2)=\tilde{\alpha}_j$. 

From (\ref{conclu1}) and (\ref{conclu3}), we have for each $j$, as $d\rightarrow 0$,
\begin{align}
&d^2N_j\longrightarrow \tilde{a}_j^2 \sigma^2\mu~~~~\text{\rm almost surely},\nonumber \\
&d^2E(N_j)\longrightarrow \tilde{a}_j^2 \sigma^2\mu, \nonumber
\end{align}
which, combined with $\tilde{a}^2_1+\tilde{a}^2_2=a^2$, show that
\begin{align}
&d^2N^*=d^2(N_1+N_2)\longrightarrow (\tilde{a}_1^2 +\tilde{a}_2^2)\sigma^2\mu=a^2\sigma^2\mu~~~~\text{\rm almost surely},\nonumber \\
&d^2E(N^*)=d^2E(N_1+N_2)\longrightarrow (\tilde{a}_1^2 +\tilde{a}_2^2) \sigma^2\mu=a^2\sigma^2\mu. \nonumber
\end{align}
It follows that
\begin{align}
&\lim_{d\rightarrow 0}\frac{d^2N^*}{a^2\sigma^2\mu}=1 ~~~~\text{\rm almost surely}, \nonumber \\
&\lim_{d\rightarrow 0}\frac{d^2E(N^*)}{a^2\sigma^2\mu}=1. \nonumber
\end{align}

As $d$ tends to 0, we know that $\sqrt{N_j}(\hvesub{\beta}{N_j}-\vesub{\beta}{0})$ has an asymptotic normal distribution and the variance of $\hvesub{\beta}{N_j}$ can be estimated with $(\vesub{X}{N_j}\vess{X}{N_j}{T})^{-1}\sigma^2$. Since the samples for the two machines are independent, the variance of $\hve{\beta}$ can be estimated with
$$ \sigma^2 \{\rho_1^2(\vesub{X}{1N_1}\vess{X}{1N_1}{\top})^{-1}+\rho_2^2(\vesub{X}{2N_2}\vess{X}{2N_2}{\top})^{-1}\}.$$
It follows that as $d\rightarrow 0$,
\begin{align}
\sigma^{-2}(\hve{\beta}-\vesub{\beta}{0})^\top \left[\left\{\rho_1^2(\vesub{X}{1N_1}\vess{X}{1N_1}{\top})^{-1}+\rho_2^2(\vesub{X}{2N_2}\vess{X}{2N_2}{\top})^{-1}\right\}\right]^{-1}(\hve{\beta}-\vesub{\beta}{0})\longrightarrow \chi_p^2.\label{dist}
\end{align}

By definition, $\mu_{N^*}=\rho_1\mu_{1N_1}+\rho_2\mu_{2N_2}$, $\mu_{jN_j}=\lambda_{min}[(\vesub{X}{jN_j}\vess{X}{jN_j}{\top})/N_j]$, $j=1,2$,
and $\mu=\lambda_{min}(\Sigma)$.
Hence, as $d\rightarrow 0$, we have $\mu_{N^*}\rightarrow \mu$ and from (\ref{Fconclu1}), ${N^*d^2}/{(\sigma^2\mu_{N^*})}\rightarrow a^2$ almost surely. Therefore, (\ref{dist}) implies that
\begin{align}
&\lim_{d\rightarrow 0} P(\vesub{\beta}{0} \in R_{N^*})\nonumber\\
=& \lim_{d\rightarrow 0} P\left((\hve{\beta}-\vesub{\beta}{0})^\top \left[\left\{\rho_1^2(\vesub{X}{1N_1}\vess{X}{1N_1}{\top})^{-1}+\rho_2^2(\vesub{X}{2N_2}\vess{X}{2N_2}{\top})^{-1}\right\}\right]^{-1}(\hve{\beta}-\vesub{\beta}{0})\leq \frac{N^*d^2}{\mu_{N^*}}\right)\nonumber\\
=& \lim_{d\rightarrow 0} P\left(\sigma^{-2}(\hve{\beta}-\vesub{\beta}{0})^\top \left[\left\{\rho_1^2(\vesub{X}{1N_1}\vess{X}{1N_1}{\top})^{-1}+\rho_2^2(\vesub{X}{2N_2}\vess{X}{2N_2}{\top})^{-1}\right\}\right]^{-1}(\hve{\beta}-\vesub{\beta}{0})\leq a^2\right)\nonumber\\
=& 1-\alpha.\nonumber
\end{align}
Therefore, Theorem 1 is proved.

\noindent{\bf Maximum axis of $R^{'}_{N^*}$}: The length of the maximum axis of the ellipsoid $R^{'}_{N^*}$ is
\begin{align}
D^{'}&= 2d\left(\frac{N^*}{\mu_{N^*}}\right)^{1/2}\lambda^{1/2}_{max}\left\{(\vesub{X}{1N_1}\vess{X}{1N_1}{\top}+\vesub{X}{2N_2}\vess{X}{2N_2}{\top})^{-1}\right\}.\nonumber
\end{align}
Denoted by $E=\vesub{X}{1N_1}\vess{X}{1N_1}{\top}$ and $F=\vesub{X}{2N_2}\vess{X}{2N_2}{\top}$.
It shows that
\begin{align}
&\lambda_{max}\left\{(\vesub{X}{1N_1}\vess{X}{1N_1}{\top}+\vesub{X}{2N_2}\vess{X}{2N_2}{\top})^{-1}\right\}=\frac{1}{\lambda_{min}\left\{E+F\right\}}\leq \frac{1}{\lambda_{min}(E)+\lambda_{min}(F)},\nonumber\\
&\mu_{N^*}=\rho_1 \mu_{N_1} +\rho_2 \mu_{N_2}=\rho_1 N_1\lambda_{max}(E^{-1}) +\rho_2 N_2 \lambda_{max}(F^{-1})=\frac{\rho_1N_1}{\lambda_{min}(E)}+\frac{\rho_2N_2}{\lambda_{min}(F)},\nonumber
\end{align}
which indicates that when $\rho_j=N_j/N^*$,
\begin{align}
\frac{1}{D^{'2}}&\geq \frac{1}{4d^2}\left(\frac{\rho_1^2}{\lambda_{min}(E)}+\frac{\rho_2^2}{\lambda_{min}(F)}\right)(\lambda_{min}(E)+\lambda_{min}(F))\nonumber\\
&=\frac{1}{4d^2}\left(\rho_1^2+\rho_2^2+\rho_1^2\frac{\lambda_{min}(F)}{\lambda_{min}(E)}+\rho_2^2\frac{\lambda_{min}(E)}{\lambda_{min}(F)}\right)\nonumber\\
&\geq \frac{1}{4d^2} \left(\rho_1^2+\rho_2^2+2\rho_1\rho_2 \right)=\frac{1}{4d^2}.\nonumber
\end{align}
Hence, $D^{'}\leq 2d$.

\subsection*{C. Proof of Theorem 2}

Under D-optimality, the new samples are selected with the maximized determinant of the information matrix. Then, from condition (A1) of Theorem 1, this implies
that the minimum eigenvalue of $\vesub{X}{jn}\vess{X}{jn}{\top}$ still has an order of $n$, where $\vesub{X}{jn}$ is the design matrix under the current status (i.e. $(\vesub{x}{j1},...,\vesub{x}{jn})_{p\times n}$). Therefore, the proof of Theorem \ref{Th2} follows arguments similar
to those for Theorem 1 and \citet[Theorem 8]{wangchang13}. Hence, the details are omitted here.

\subsection*{D. Properties of distributed sequential estimation with adaptive shrinkage estimation}

\noindent{\bf Partition of design matrix}: By simple computation, we have that
\begin{align*}
  &(\tve{\beta}-\ve{Z})^\top \left[\vesup{I}{*}\left\{\sum_{j=1}^M \vesub{X}{jN_j}\vess{X}{jN_j}{\top}\right\}^{-1}\vesup{I}{*}\right]^{-}(\tve{\beta}-\ve{Z})\\
  &=(\ve{O}\tve{\beta}-\ve{O}\ve{Z})^\top \left[\ve{O}\vesup{I}{*}\vesup{O}{\top}\left\{\sum_{j=1}^M \ve{O}\vesub{X}{jN_j}(\ve{O}\vesub{X}{jN_j})^{\top}\right\}^{-1}\ve{O}\vesup{I}{*}\vesup{O}{\top}\right]^{-}(\ve{O}\tve{\beta}-\ve{O}\ve{Z}).
\end{align*}
According to $\tvesub{\beta}{1}$ and $\tvesub{\beta}{2}$,
partition the matrix $\sum_{j=1}^M \ve{O}\vesub{X}{jN_j}(\ve{O}\vesub{X}{jN_j})^{\top}$ as follows,
\begin{align}
\sum_{j=1}^M \ve{O}\vesub{X}{jN_j}(\ve{O}\vesub{X}{jN_j})^{\top}=\left(
  \begin{array}{ll}
    \vesub{\Sigma}{11} & \vesub{\Sigma}{12} \\
    \vess{\Sigma}{12}{\top} & \vesub{\Sigma}{22} \\
  \end{array}
\right),\nonumber
\end{align}
where $\vesub{\Sigma}{11}$, $\vesub{\Sigma}{12}$ and $\vesub{\Sigma}{22}$ are ${\hat p_0 \times \hat p_0}$,
${\hat p_0 \times (p-\hat p_0)}$ and ${(p-\hat p_0) \times (p-\hat p_0)}$ matrices, respectively.
Then, we have
\begin{align}\label{parmat}
\ve{O}\vesup{I}{*}\vesup{O}{\top}\left\{\sum_{j=1}^M \ve{O}\vesub{X}{jN_j}(\ve{O}\vesub{X}{jN_j})^{\top}\right\}^{-1}\ve{O}\vesup{I}{*}\vesup{O}{\top}
=\left(
  \begin{array}{cc}
    \tvess{\Sigma}{11}{-1}& 0 \\
    0 & 0 \\
  \end{array}
\right),
\end{align}
where
\begin{align}\label{matrix}
\tvesub{\Sigma}{11}=\vesub{\Sigma}{11}-\vesub{\Sigma}{12}\vess{\Sigma}{22}{-1}\vess{\Sigma}{12}{\top}.
\end{align}
Hence,
\begin{align}
\tve{\Sigma}=\left(
  \begin{array}{cc}
    \tvesub{\Sigma}{11}& 0 \\
    0 & 0 \\
  \end{array}
\right)\nonumber
\end{align}
is a general inverse matrix of the matrix defined on the left-hand side of (\ref{parmat}).
Consequently, we have
\begin{align*}
&\hat\sigma^{-2}(\tve{\beta}-\vesub{\beta}{0})^\top \left[\vesup{I}{*}\left\{\sum_{j=1}^M \vesub{X}{jN_j}\vess{X}{jN_j}{\top}\right\}^{-1}\vesup{I}{*}\right]^{-}(\tve{\beta}-\vesub{\beta}{0})\nonumber\\
& = \hat\sigma^{-2}(\tvesub{\beta}{1}-\vesub{\beta}{01})^\top \tvesub{\Sigma}{11}(\tvesub{\beta}{1}-\vesub{\beta}{01}).
\end{align*}

\noindent{\bf Proof of Theorem 3}: At the stopping time $\tilde{N}_j$, similar to Lemma 4, we have
\begin{align}
&\lim_{d\rightarrow 0}\frac{d^2\tilde{N}_j}{\tilde{a}_j^2\sigma^2\tilde{\mu}}=1 ~~~~\text{\rm almost surely}, \nonumber \\
&\lim_{d\rightarrow 0}\frac{d^2E(\tilde{N}_j)}{\tilde{a}_j^2\sigma^2\tilde{\mu}}=1, \nonumber
\end{align}
which indicates that for each $j$, as $d\rightarrow 0$,
\begin{align}
&d^2\tilde{N}_j\longrightarrow \tilde{a}_j^2 \sigma^2\tilde{\mu}~~~~\text{\rm almost surely},\nonumber \\
&d^2E(\tilde{N}_j)\longrightarrow \tilde{a}_j^2 \sigma^2\tilde{\mu}. \nonumber
\end{align}
Hence, from $\tilde{a}^2_1+\tilde{a}^2_2=a^2$, it shows that
\begin{align}
&\lim_{d\rightarrow 0}\frac{d^2\tilde{N}}{a^2\sigma^2\tilde{\mu}}=1 ~~~~\text{\rm almost surely}, \nonumber \\
&\lim_{d\rightarrow 0}\frac{d^2E(\tilde{N})}{a^2\sigma^2\tilde{\mu}}=1. \nonumber
\end{align}

From \cite{wangchang13}, we have as $d\rightarrow 0$,
$$\vesub{I}{j}(\epsilon)\longrightarrow \vesub{I}{0}~~ \text{\rm almost surely}, $$
where $\vesub{I}{0}={\rm{diag}}\{I(\beta_{01}\neq 0),\cdots,\beta_{0p}\neq 0)\}$. Since the sequential sampling procedures for $M$ machines are independent, as $d\rightarrow 0$,
$$\vesup{I}{*}=\prod_{j=1}^M \vesub{I}{j}(\epsilon)\longrightarrow \vesub{I}{0}~~ \text{\rm almost surely},$$
which gives that
$$\lim_{d\rightarrow 0}\hat{p}_0= p_0~~ \text{\rm almost surely}.$$
Furthermore, it is easily shown that $\lim_{d\rightarrow 0}E(\hat{p}_0)= p_0$. Similar to proof of Theorem 1, we have
 $$\lim_{d\rightarrow 0} P(\vesub{\beta}{0} \in R_{\tilde{N}})=1-\alpha.$$
The Proof of Theorem 3 is finished.

\bibliographystyle{elsarticle-harv}

\end{document}